\documentclass[epsfig,12pt]{article}

\usepackage{epsfig,amssymb,euscript}
\usepackage{amsmath,amscd}

\numberwithin{equation}{section}

\newcommand{\ii}{\mathrm{i}}
\newcommand{\dd}{\mathrm{d}}
\newcommand{\me}{\mathrm{e}}

\newcommand{\vol}{\mathrm{vol}}

\DeclareMathOperator{\Cliff}{Cliff}
\DeclareMathOperator{\Spin}{\mathit{Spin}}

\newcommand{\be}{\begin{eqnarray}} \newcommand{\ee}{\end{eqnarray}}
\newcommand{\bea}{\begin{eqnarray}} \newcommand{\eea}{\end{eqnarray}}
\newcommand{\ba}{\begin{array}} \newcommand{\ea}{\end{array}}

\newcommand{\nn}{\nonumber \\}

\newcommand{\p}[1]{(\ref{#1})}

\def\bbox{{\,\lower0.9pt\vbox{\hrule \hbox{\vrule height 0.2 cm \hskip
0.2 cm \vrule height 0.2 cm}\hrule}\,}}

\newcommand{\dsl}{\pa \kern-0.5em /}

\newcommand{\diag}{{\rm diag}\,}

\newcommand{\trsp}{\mathrm{T}}

\newcommand{\de}{\partial} \newcommand{\diff}{\mathrm{d}}

\def\ds{\raise.15ex\hbox{/}\kern-.57em\partial}
\def\Ds{\,\raise.15ex\hbox{/}\mkern-13.5mu D}

\renewcommand{\a}{\alpha} \renewcommand{\b}{\beta}
 
\newcommand{\e}{\epsilon}

\newcommand{\bCP}{\mathbb{C}P}
\newcommand{\bR}{\mathbb{R}}
\newcommand{\bC}{\mathbb{C}}
\newcommand{\bP}{P}
\newcommand{\bZ}{\mathbb{Z}}

\def\a{\alpha}\def\b{\beta}
\def\s{\sigma}

\begin{document}


\begin{titlepage}

\vfill

\begin{flushright}
Imperial/TP/03-04/6\\ hep-th/0402153\\
\end{flushright}

\vfill

\begin{center}

\baselineskip=16pt

{\Large\bf Supersymmetric $AdS_5$ solutions of M-theory}

\vskip 1.3cm

Jerome P. Gauntlett$^{1*}$, Dario Martelli$^{2}$, James Sparks$^{2}$
and Daniel Waldram$^{2}$

\vskip 1cm

{\small{\it $^1$Perimeter Institute for Theoretical
Physics\\ Waterloo, ON, N2J 2W9, Canada\\ E-mail:
jgauntlett@perimeterinstitute.ca\\}}
\vskip .6cm
{\small{\it
$^{2}$Blackett Laboratory, Imperial College\\ London, SW7 2BZ, U.K.\\
E-mail: d.martelli, j.sparks, d.waldram@imperial.ac.uk \\}}

\end{center}

\vfill

\begin{center}
\textbf{Abstract}
\end{center}

\begin{quote}

We analyse the most general supersymmetric solutions of $D=11$
supergravity consisting of a warped product of five-dimensional
anti-de-Sitter space with a six-dimensional Riemannian space
$M_6$, with four-form flux on $M_6$. We show that $M_6$ is partly
specified by a one-parameter family of four-dimensional K\"ahler
metrics. We find a large family of new explicit regular solutions
where $M_6$ is a compact, complex manifold which is topologically
a two-sphere bundle over a four-dimensional base, where the latter
is either (i) K\"ahler-Einstein with positive curvature,
or (ii) a product of two constant-curvature Riemann surfaces.
After dimensional reduction and T-duality,
some solutions in the second class are related to a new
family of Sasaki-Einstein spaces which includes $T^{1,1}/\bZ_2$.
Our general analysis also covers warped products of five-dimensional
Minkowski space with a six-dimensional Riemannian space.

\end{quote}

\vfill

\vfill \vskip 5mm \hrule width 5.cm \vskip 5mm
{\small{\noindent $^*$ On leave from: Blackett Laboratory, Imperial
  College, London, SW7 2BZ, U.K.\\
  }}

\end{titlepage}


\section{Introduction}

String or M-theory on a supersymmetric background that contains an
$AdS_5$ factor is expected to be equivalent to a four-dimensional
superconformal field theory~\cite{mal}.  The best-known example is
the $AdS_5\times S^5$ solution of type IIB string theory which is
conjectured to be dual to $N=4$ super-Yang-Mills theory. This
geometry arises as the near-horizon limit of the supergravity
solution describing D3-branes in flat space. It is a special case
of a general class of supersymmetric solutions of the form
$AdS_5\times M_5$ where $M_5$ is a Sasaki-Einstein five-manifold.
These arise as the near-horizon limits of solutions describing
D3-branes at the singularities of Calabi--Yau cones and are dual to
$N=1$ superconformal field
theories~\cite{Klebanov:1998hh,Acharya:1998db,Morrison:1998cs}.

The purpose of this paper is primarily to study analogous
solutions in M-theory. We first analyse the most general class of
supersymmetric solutions of the form of a warped $AdS_5\times M_6$
product, deriving the conditions on the geometry of the six-dimensional
manifold $M_6$. We then use our results to construct
rich new families of explicit regular compact solutions.

To date, rather surprisingly, only a handful of supersymmetric
$AdS_5$ solutions have been found in M-theory (There are also some
examples of non-supersymmetric
solutions~\cite{Pope:1988xj,Gauntlett:2002rv}). A rather trivial
case is the maximally supersymmetric $AdS_7\times S^4$ solution:
$AdS_7$ can be foliated by $AdS_5$ and $S^1$ (see, for example,
ref.~\cite{Cvetic:2000cj}) and hence this solution can be viewed
as a warped product of $AdS_5$ with a (non-compact)
six-dimensional space. Another example was presented in
ref.~\cite{Alishahiha:1999ds} where it was interpreted as the
near-horizon limit of two semi-localised M5-branes (the same
solution also appears in ref.~\cite{Fayyazuddin:1999zu}). It was
subsequently shown in ref.~\cite{Cvetic:2000cj} that this solution
can be obtained from the type IIB $AdS_5\times S^5$ solutions by a
simple T-duality followed by an uplift to $D=11$. It was noted in
ref.~\cite{Alishahiha:1999ds} that the six-dimensional manifold is
singular indicating that the solution is not capturing all of the
degrees of freedom of the intersecting M5-brane system, and hence
the $D=11$ solution is of limited utility in analysing the dual
field theory.

Two more interesting examples were found in ref.~\cite{malnun}.
These solutions describe the near horizon limit of fivebranes
wrapping holomorphic curves in Calabi--Yau two- or three-folds. In
these solutions $M_6$ is an $S^4$ bundle over $H^2/\Gamma$ where
$\Gamma$ is a discrete group of isometries of $H^2$, and hence
$H^2/\Gamma$ can be compact. They are dual to the $N=2$ or $N=1$
superconformal field theories arising on fivebranes wrapped on a
holomorphic $H^2/\Gamma$ cycle in a Calabi--Yau two- or
three-fold, respectively. These solutions were first found in
$D=7$ gauged supergravity and then uplifted to obtain solutions of
$D=11$ supergravity. Regular and compact solutions where $M_6$ is
an $S^4$ bundle over $S^2$ were found in ref.~\cite{Cucu:2003bm}
but these have yet to be connected with a dual field theory.
Finally, $AdS_5$ solutions were also found in
ref.~\cite{Fayyazuddin:2000em}, and were argued to be related to
$N=2$ superconformal field theories arising on intersecting
fivebranes. It would be interesting to know if these solutions are
regular.

This small collection of solutions have all been found by guessing
a suitable ansatz for the metric on $M_6$. Here we will systematically
determine the general conditions placed on the geometry $M_6$, the four-form
field strength and the warp factor in order for the solution to have
$N=1$ supersymmetry. These conditions thus characterise the
most general way in which four-dimensional superconformal field
theories can arise in M-theory via the AdS/CFT correspondence.
Furthermore,
we use our
results to obtain a rich family of new solutions in explicit form.
It is quite simple to extend our analysis to characterise the most
general warped products of five-dimensional Minkowski space with a
six-dimensional manifold and this analysis is included in an appendix.

Our method follows that employed to analyse the most general kinds
of supersymmetric solutions in supergravity theories, using the
language of ``$G$-structures'' and ``intrinsic torsion''. It
provides a systematic way of translating the local supersymmetry
constraints into differential conditions on a set of differential forms on
$M_6$ defining the metric and flux. The utility of $G$-structures
for analysing supersymmetric solutions was first advocated in
ref.~\cite{Gauntlett:2002sc}; for related developments in various
string/M-theory settings see
refs.~\cite{Gurrieri:2002wz}--\cite{Ivanov:2003nd}. Essentially
the same techniques have also proved very useful in determining
the general form of supersymmetric solutions in various
lower-dimensional
supergravities~\cite{Gauntlett:2002nw}--\cite{Cariglia:2004kk}. In
the present setting, we first concentrate on the local conditions
on the geometry. We will see that the Killing spinor defines a
preferred local $SU(2)$ structure on $M_6$, characterised by a
number of tensor fields constructed as bilinears in the Killing
spinor. The Killing spinor equation then implies a number of
differential conditions on the tensors that determine the
intrinsic torsion of the structure.

It is then straightforward to see what these conditions imply for
the geometry of $M_6$. We will show that the metric on $M_6$
always admits a Killing vector. In terms of the $AdS$/CFT
correspondence this corresponds to the R-symmetry of the $N=1$
superconformal field theory. Locally the five-dimensional space
orthogonal to the Killing vector is a warped product of a
one-dimensional space with coordinate $y$ and a four-dimensional
complex space with a one-parameter family of K\"ahler metrics
depending on $y$.

We then use these results to construct new explicit solutions. To
do this we impose a very natural geometric condition on the
geometry. Specifically, we demand that the six-dimensional space
is a complex manifold. Globally, the new regular compact solutions
that we construct are all holomorphic two-sphere bundles over a
smooth four-dimensional K\"ahler base $M_4$. Using a recent
mathematical result on K\"ahler manifolds~\cite{apostolov}, we are
able to completely classify this class of solutions (assuming that
the Goldberg conjecture is true).  In particular, at fixed $y$ the
base is either (i) a K\"ahler--Einstein (KE) space or (ii) a
non-Einstein space which is the product of two constant curvature
Riemann surfaces.

In the KE class (i) we find regular solutions only if the
curvature is positive. Such spaces have been classified by Tian
and Yau~\cite{tian,tianyau} and are topologically either
$S^2\times S^2$, $\mathbb{C}P^2$ or $\mathbb{C}P^2\#_n
\mathbb{C}P^2$ with $n=3,\dots,8$. The KE metrics are of course
explicitly known in the first two cases, but are not known
explicitly for the other examples. In the second product class
(ii) one finds regular compact solutions when the geometry is
$S^2\times S^2$, or $S^2\times T^2$. There are also regular
solutions of the form $S^2\times H^2$, where $H^2$ is hyperbolic
space, but these are not compact. Note, however, that we recover
the $N=1$ solution of ref.~\cite{malnun}, for which it is known
that one can divide $H^2$ by a discrete group of isometries to
obtain compact solutions while preserving supersymmetry. 
A number of additional singular
solutions in both the first and second classes are also found in
explicit form.

The $S^2\times T^2$ class of solutions have the particularly
interesting property that they can be reduced on an $S^1$ in $T^2$
to obtain regular supersymmetric type IIA supergravity solutions
of the form $AdS_5\times X_5'$. Then, via a T-duality on the other
$S^1$, these give solutions of the form $AdS_5 \times X_5$ where
$X_5$ is a new one-parameter family of Sasaki-Einstein spaces.
Global properties of these will be studied in more detail in a
separate paper \cite{paper2}, but we note that one special case
gives $AdS_5\times T^{1,1}/\bZ_2$, where $T^{1,1}/\bZ_2$ is a
well-known Sasaki--Einstein manifold\footnote{Note here, and
throughout the paper, $T^{1,1}/\bZ_2$
  refers to the unique smooth quotient of $T^{1,1}$ by $\bZ_2$ that preserves the Sasaki-Einstein
  structure.}; the superconformal field theory was identified in
ref.~\cite{Morrison:1998cs} (see also
ref.~\cite{Klebanov:1998hh}). It will clearly be interesting to
identify dual conformal field theories for our new solutions and
determine under which conditions the M-theory, type IIA or type
IIB supergravity solution is most useful.

The plan of the rest of the paper is as follows.
Section~\ref{susy-conds} contains the analysis of the conditions
imposed on the geometry of $M_6$ in order to get a supersymmetric
solution. We have tried to minimise the details as much as
possible, relegating most of the calculation to two
appendices~\ref{conds} and~\ref{su2structure}. The conditions on
the local form of the metric and flux are summarized at the end of
section~\ref{introducingcoo}. Section~\ref{special} starts by
analysing the additional local conditions on the geometry that
arise by assuming that $M_6$ is a complex manifold. We then
discuss a natural ansatz for the global topology of the solutions,
requiring $M_6$ to be an $S^2$ fibration over a four-dimensional
manifold $M_4$. Regular compact solutions of this type are shown
to fall into two classes: one where the base is KE, the other
where it is a product of constant curvature Riemann surfaces. In
either case we show that the supersymmetry conditions reduce to
solving a single non-linear ordinary differential equation.
Section~\ref{sec:KE} discusses the explicit solutions in the first
class, for positive, negative and zero curvature.
Section~\ref{sec:product} discusses the explicit solutions of the
second type, again for each different possible sign of curvature.
We also discuss the type IIA and IIB duals to the $S^2\times T^2$
solutions and how the solution of ref.~\cite{malnun} is obtained
in our formalism. Finally, we end with some conclusions. In the
remaining appendices we give a summary of our conventions
(appendix~\ref{app:conv}) and analyse the closely related case of
supersymmetric warped products of Minkowski space with $M_6$
(appendix~\ref{m=0}).


\section{The conditions for supersymmetry}
\label{susy-conds}

We want to find the general structure of supersymmetric configurations
of $D=11$ supergravity that are warped products of an external $AdS_5$
space with an internal six-dimensional Riemannian manifold $M_6$:
\bea
\label{ansatz}
   \dd s^2&=&\me^{2\lambda(v)}[\dd s^2(AdS_5)+\dd s^2(M_6)]\nn
   G&=&\frac{1}{4!}G_{\mu_1\mu_2\mu_3\mu_4}\dd v^{\mu_1}\wedge \dd v^{\mu_2}
      \wedge \dd v^{\mu_3}\wedge \dd v^{\mu_4}
\eea
where $u^\alpha$, $\alpha=0,1,\dots, 4$ are co-ordinates on $AdS_5$
and $v^\mu$, $\mu=1,2,\dots 6$ are co-ordinates on $M_6$.  We assume
that the $AdS_5$ space has radius squared given by $m^{-2}$ so that
\begin{equation}
   R_{\alpha\beta}=-4m^2g_{\alpha\beta}~.
\end{equation}
Note that when $m=0$ our ansatz reduces to warped products of
five-dimensional Minkowski space with $M_6$. For completeness, this
case is discussed separately in appendix~\ref{m=0}.

Our conventions for $D=11$ supergravity are included in
appendix~\ref{app:conv}.
We want solutions admitting at least one supersymmetry and satisfying
the equations of motion. Supersymmetry implies we have a solution of
the Killing spinor equation
\begin{equation}
\label{ks}
   \nabla_\mu\e + \frac{1}{288}\left[
         \Gamma{_\mu}{^{\nu_1\nu_2\nu_3\nu_4}}
         - 8\delta{_\mu^{\nu_1}}\Gamma^{\nu_2\nu_3\nu_4}
      \right]G_{\nu_1\nu_2\nu_3\nu_4}\e
      = 0
\end{equation}
where $\epsilon$ is a Majorana spinor in the representation where
$\Gamma_{11}\equiv\Gamma_0\Gamma_1\dots\Gamma_{10}=1$. To see how this
reduces to a condition on $M_6$ we first decompose the $D=11$ Clifford
algebra $\mathrm{Cliff}(10,1)\cong
\mathrm{Cliff}(1,4)\otimes \mathrm{Cliff}(6,0)$.  Explicitly,
$\mathrm{Cliff}(6,0)\cong H(4)$ and $\mathrm{Cliff}(1,4)\cong
H(2)\oplus H(2)$ and hence the tensor product gives $R(32)\oplus
R(32)\cong\mathrm{Cliff}(10,1)$. In other words, we can write the
$D=11$ gamma matrices as
\bea
   \Gamma^a&=&\rho^a\otimes \gamma_7\nn
   \Gamma^m&=&1\otimes \gamma_m
\eea
where $a,b=0,1,\dots,4$ and $m,n=1,2,\dots,6$ are frame indices on
$AdS_5$ and $M_6$ respectively, and we have
\be
   [{\rho^a},{\rho^b}]_+ = -2\eta^{ab},\qquad
   [{\gamma^m},{\gamma^n}]_+ =2\delta^{mn}
\ee
with $\eta^{ab}=\diag(-1,1,1,1,1)$. Note that
\be
   \gamma_7\equiv \gamma_1\dots\gamma_6
\ee
so that $(\gamma_7)^2=-1$. Given our conventions in $D=11$, we then
have $\rho_{01234}=-1$. The parameter $\epsilon$ decomposes as
$\psi(u)\otimes e^{\lambda/2}\xi(v)$, where the dependence on
$\lambda$ is added to simplify the resulting formulae. On the $AdS_5$
space, the Killing spinor satisfies
\begin{equation}
   D_a\psi \equiv \left(
      \partial_a - \frac{1}{4}\omega_{abc}\rho^{bc}\right)\psi
      = \frac{1}{2}im\rho_a\psi~.
\end{equation}
The $D=11$ Killing spinor equation then implies that the internal
$\xi$ must satisfy
\bea\label{susycondsone}
   \left[\nabla_m+\frac{1}{2}im\gamma_m\gamma_7 -
      \frac{1}{24}\me^{-3\lambda}\gamma^{n_1n_2n_3}
         G_{mn_1n_2n_3}\right]\xi &=& 0~{} \nn
   \left[\gamma^m\nabla_m\lambda+\frac{1}{144}
      \me^{-3\lambda}\gamma^{m_1m_2m_3m_4}G_{m_1m_2m_3m_4}
      - im \gamma_7\right]\xi &=& 0~.
\eea

For a true supergravity solution we must check that a background
satisfying these equations is actually a solution of the equations
of motion. By analysing the integrability conditions, and using
the arguments presented in ref.~\cite{Gauntlett:2002fz}, we find
that a geometry admitting solutions to~\eqref{susycondsone} will
satisfy the equations of motion if the Bianchi identity and the
equation of motion for the four-form $G$ are imposed. Given our
ansatz~\eqref{ansatz} these reduce to the following two conditions
on $M_6$:
\begin{equation}
\label{eqmot}
   \dd(\me^{3\lambda}*_6G) = 0
   \qquad \dd G = 0~.
\end{equation}

The spinor $\xi$ is a representation of the Clifford algebra
$\Cliff(6,0)$. Under $\Spin(6,0)$ it decomposes into chiral and
anti-chiral pieces $\xi=\xi_++\xi_-$ where
$-\ii\gamma_7\xi_\pm=\pm \xi_\pm$. It is straightforward to
analyse the Killing spinor equations~\eqref{susycondsone} in the
particular case that $\xi$ is chiral. For either chirality, one
finds that $m=0$, $\lambda=\text{constant}$, $G=0$ and
$\nabla\xi=0$.  This implies that solutions are, after possibly
going to the covering space, simply the direct product
$\bR_{1,4}\times M_6$, with $M_6$ a Calabi--Yau threefold, and
with zero $G$-flux. In particular, there are no $AdS_5$ solutions,
which is the case of primary interest in this paper\footnote{Note
that this is somewhat analogous to M-theory compactifications on
eight-manifolds, where it was shown in
ref.~\cite{Becker:1996gj,MS} that a chiral internal spinor rules
out most of the possible internal fluxes, and fixes the internal
space to be conformal to a $Spin(7)$-holonomy manifold.}.
Henceforth we therefore assume that $\xi$ is of indefinite
chirality.

In summary, we want to analyse~\eqref{susycondsone}, combined with
the Bianchi identity and equation of motion~\eqref{eqmot} for $G$.
In particular, we will recast these conditions into equivalent
conditions on the local geometry of $M_6$, the warp factor and the
four-form.

\subsection{Spinor bilinears and local $SU(2)$-structure}

We will derive the local form of the metric by first deriving a
set of differential conditions on a set of spinor bilinears formed
from $\xi$. The details of the calculation are described in
appendices~\ref{conds} and~\ref{su2structure}. First note that
generically the non-chiral $\xi$ can be decomposed as 
\be
   \xi\equiv\xi_+ + \xi_-=f_1\eta_1 +f_2[a\eta_1 +(1-|a|^2)^{1/2}\eta_2]^*
\ee
where $\eta_i$ are two orthogonal unit-norm chiral spinors, and
$f_i$ and $a$ are real and complex functions, respectively. 
As we show in appendices~\ref{conds}
and~\ref{su2structure}, the supersymmetry
conditions~\eqref{susycondsone} imply that $\xi$ has constant norm
and furthermore $\xi^\trsp\xi=0$. This implies we can choose a
normalization such that we have
\begin{equation}
   \xi = {\sqrt 2}(\cos\alpha\eta_1 + \sin\alpha\eta_2^*)~.
\end{equation}

Since the stabilizer of $(\eta_1,\eta_2)$ in $SO(6)$ is $SU(2)$
they define a particular privileged local $SU(2)$ structure on
$M_6$. Equivalently the structure can be specified by a set of
bilinear forms constructed from $\eta_i$. It is worth emphasizing
that since $\eta_i$ are not globally defined in general (in
particular, one or the other is not defined when $\sin\alpha=0$ and
$\cos\alpha=0$), the $SU(2)$ structure on $M_6$ is not globally
defined in general either. We will address the global
$G$-structure in the case of a specific class of global solutions
in section~\ref{special}.

Nevertheless, in deriving the local form of the metric it is still
very useful to derive differential conditions on the local $SU(2)$
structure.  A basic set of bilinears specifying the local $SU(2)$
structure is given in  equations~\p{defplus} and~\p{defplustwo} in
appendix~\ref{su2structure}. They consist of a fundamental
$(1,1)$-form $J$, a  complex $(2,0)$-form $\Omega$ and two
one-forms $K^1$ and $K^2$. The metric can be written as
\be
   \diff s^2=e^ie^i+(K^1)^2+(K^2)^2
\ee
where $J=e^1\wedge e^2+e^3\wedge e^4$ and $\Omega=(e^1+\ii
e^2)\wedge(e^3+\ii e^4)$. The volume form is defined by
$\vol_6=\frac{1}{2}J\wedge J \wedge K^1\wedge K^2$.  These
forms satisfy the following set of differential constraints,
derived in appendices~\ref{conds} and~\ref{su2structure}:
\bea
   \me^{-3\lambda}\diff(\me^{3\lambda}\sin\zeta) & = & 2 m
      K^1\cos\zeta\label{ns.a}\\
   \me^{-6\lambda}\diff (\me^{6\lambda} \Omega\cos\zeta ) & = & 3 m \, \Omega
      \wedge (-K^1 \sin\zeta + i  K^2)\label{ns.c}\\
   \me^{-6\lambda}\diff(\me^{6\lambda} K^2 \cos\zeta) & = &
      \me^{-3\lambda}* G + 4 m (J - K^1\wedge K^2\,\sin\zeta)
      \label{ns.d}\\
   \me^{-6\lambda}\diff (\me^{6\lambda}J \wedge K^2 \cos\zeta) & = &
      \me^{-3\lambda}G \sin\zeta + m  (J \wedge J
      - 2 J \wedge K^1 \wedge K^2 \sin\zeta )~.\label{ns.e}
\eea
In the last formula we have used $*J=J\wedge K^1 \wedge K^2$. Here
we have defined $\zeta=\pi/2-2\alpha$, giving
$\cos2\alpha=\sin\zeta$, $\sin2\alpha=\cos\zeta$. Note that
multiplying each of the first three equations by a suitable power
of $\me^\lambda$ and taking the exterior derivative so that the
left-hand side vanishes leads to three more equations which should be
separately imposed when $m=0$. (This case is discussed separately
in appendix~\ref{m=0}.)

We will argue in the following section that the above differential
conditions are the set of necessary and sufficient local
differential conditions for the geometry to admit a Killing
spinor. In addition we will show that they automatically imply
that the equation of motion~\eqref{eqmot} and (for $m\ne 0$) the
Bianchi identity for $G$ are satisfied.

\subsection{Reduction from $d=7$}

Before analysing what these conditions imply about the local form of
the metric, we briefly pause to discuss how they can be obtained in a
different way.  If we write $AdS_5$ in Poincar\'e co-ordinates, the
warped product of $AdS_5$ with $M_6$ that we are considering can be
viewed as a special case of a warped product of Minkowski four-space
$\bR^{1,3}$ with a seven-dimensional Riemannian manifold $M_7$. That
is, we can rewrite~\eqref{ansatz} as
\begin{equation}
\label{M7}
   \dd s^2  = \me^{2\lambda} \me^{-2mr}\dd s^2 (\bR^{1,3})
      + \me^{2\lambda} \left[ \dd r^2 + \dd s^2_6 \right]
\end{equation}
with $G$ a four-form on $M_6$ that is independent of $r$. 
General conditions on the geometry of $M_7$ imposed by
supersymmetry  have been analysed in ref.~\cite{kmt} and then
subsequently in ref.~\cite{Behrndt:2003zg,Dall'Agata:2003ir}. 
We now show how to obtain our conditions from these results.

It was shown in ref.~\cite{kmt} that the $d=7$ geometry is determined by an
$SU(3)$ structure specified by a vector $K'$, a two-form $J'$ and a
three-form $\Omega'$. In particular, we write the $d=7$ metric as
\begin{equation}
   \dd s_7^2 = e'^ae'^a + (K')^2
\end{equation}
with $J'=e'^1\wedge e'^2+e'^3\wedge e'^4+e'^5\wedge e'^6$ and
$\Omega'=(e'^1+\ii e'^2)\wedge(e'^3+\ii e'^4)\wedge(e'^5+\ii e'^6)$.
The volume form is given by
$\vol_7=\frac{1}{3!}J'\wedge J'\wedge J'\wedge K'$.  The differential
conditions are given, in our conventions, by
\begin{equation}
\label{red-conds}
\begin{aligned}
   \diff (\me^{2\Delta} K') &= 0 \\
   \me^{-4\Delta} \diff (\me^{4\Delta} J') &= *_7 G \\
   \diff(\me^{3\Delta}\Omega') &= 0 \\
   \me^{-2\Delta}\diff (\me^{2\Delta}J'\wedge J') &=
      - 2 \, G \wedge K'
\end{aligned}
\end{equation}
with metric
\be
   \dd s^2 = \me^{2\Delta} \dd s^2 (\bR^{1,3}) + \dd s^2_7~.
\ee
The first three equations were derived in ref.~\cite{kmt} while the
last equation corrects that appearing in ref.~\cite{kmt} by a factor.
It was argued in ref.~\cite{Dall'Agata:2003ir} that these are 
sufficient conditions for a geometry to 
admit a Killing spinor. Furthermore, the second equation implies the $G$
equation of motion and thus, given an integrability argument as
in ref.~\cite{Gauntlett:2002fz}, only the Bianchi identity $\dd G$ need
be imposed to give a solution to the full equations of motion.

It is easy to show that our set of equations \p{ns.a}--\p{ns.e} for the $d=6$
geometry are precisely equivalent to the conditions~\eqref{red-conds}
together with the $AdS_5$ metric ansatz~\eqref{M7}. The correspondence
is given by first identifying
$\me^{2\Delta}=\me^{2\lambda}\me^{-2mr}$. The forms $(K',J',\Omega')$
are then related to $(J,\Omega,K_1,K_2)$ by
\begin{equation}
\label{reduction}
\begin{aligned}
   K' &= \me^\lambda ( \cos \zeta K^1 -\sin \zeta \diff r ) \\
   J' &= \me^{2\lambda} (J - (\sin \zeta K^1 + \cos\zeta \diff r )
      \wedge K^2) \\
   \Omega' &= \me^{3\lambda}\,\Omega
      \wedge (-\sin\zeta K^1  - \cos\zeta \diff r  + \ii K^2)~.
\end{aligned}
\end{equation}
This structure arises as follows. In order to obtain $AdS_5$
we need to select a radial direction from the $d=7$ geometry, which we
denote by the unit one-form $\me^\lambda\diff r$.  In general this
radial direction will point partly in the $K'$ direction and partly in
the $d=6$ space orthogonal to $K'$. In other words we have
\begin{equation}
\label{K1def}
\begin{aligned}
   \me^\lambda\dd r &= -\sin\zeta K' - \cos\zeta W \\
   \me^\lambda K^1 &= \cos\zeta K' - \sin\zeta W
\end{aligned}
\end{equation}
where $W$ is a unit one-form in $M_6$. We have also defined $K^1$
as the unit one-form, orthogonal linear combination of $K'$ and
$W$. (The angle $\zeta$ is chosen to match the definition in the
previous section). The almost complex structure determined by
$J'_a{}^b$ in $d=6$ pairs the one-form $W$ with another unit
one-form $\me^\lambda K^2\equiv J'\cdot W$. With these definitions,
inverting the rotation~\eqref{K1def} to write $(K',W)$ in terms of
$(\dd r,K^1)$, one then gets the expressions in~\eqref{reduction} for
$K'$, $J'$ and $\Omega'$. 

As mentioned the $d=7$ conditions~\eqref{red-conds} are necessary
and sufficient for a geometry which admits a Killing spinor. Given
our metric ansatz they are equivalent to our
conditions~\p{ns.a}--\p{ns.e} on $M_6$. Hence our conditions are
also necessary and sufficient for supersymmetry.

To ensure we have a solution of the equations of motion, in
general one also needs to impose the equation of motion and
Bianchi identity~\eqref{eqmot} for $G$. The connection with the
$d=7$ results gives us a quick way of seeing that, in fact,
provided $\sin\zeta$ is not identically zero, both conditions are
a consequence of the supersymmetry constraints~\p{ns.a}--\p{ns.e}.
As already noted, the equation of motion for $G$ follows directly from the exterior
derivative of the second equation in~\eqref{red-conds}. For the
Bianchi identity one notes that, given the ansatz for the $d=7$
metric and $G$, the first and last equations in~\eqref{red-conds}
imply in general that
\begin{equation}
   \sin\zeta \dd G \wedge \dd r = 0
\end{equation}
since $\dd G$ lies solely in $M_6$.
This implies that $\dd G=0$ provided $\sin\zeta$ is not
identically zero -- which can only occur only when $m=0$. Thus we
see that, when $m\neq 0$, the
constraints~\eqref{ns.a}--\eqref{ns.e} are necessary and
sufficient both for supersymmetry and for a solution of the
equations of motion.

\subsection{Local form of the metric}
\label{introducingcoo}

In this section, we use the differential conditions on the forms
derived in section 2.1 to give the local form of the
metric. We start by considering $K^1$ and $K^2$.

For $K^1$, we can immediately integrate the condition~\eqref{ns.a} and
introduce coordinates $(w^M,y)$ with $M=1,\dots,5$ and $y$
defined by
\begin{equation}
   2m y = \me^{3\lambda} \sin\zeta
\end{equation}
so that
\begin{equation}
\label{K1sol}
   K^1 = \me^{-3\lambda}\sec\zeta \dd y~.
\end{equation}
While we could eliminate either $\lambda$ or $\zeta$ from the following
formulae, for the moment it will be more convenient to keep both.
The metric then has the form
\be
   \dd s^2  =  g^5_{MN} (w,y) \dd w^M \dd w^N
      + \me^{-6\lambda(w,y)}\sec^2\zeta(w,y)\dd y^2~.
\ee

Turning to $K^2$, as discussed in appendix~\ref{conds}, it is easy
to show, starting from the Killing spinor equations,
that
$\tilde{K}^2_\mu=\frac{1}{2}\bar{\xi}\gamma_\mu\gamma_7\xi=\cos\zeta
K^2_\mu$ satisfies $\nabla_{(\mu}\tilde{K}^2_{\nu)}=0$ and hence,
raising an index, we have the important condition that 
\begin{equation}
   \text{$\tilde{K}^2=\cos\zeta\, K^2\quad$ defines a Killing vector}
\end{equation}
(Note that one could also show this from the
conditions~\eqref{ns.a}--\eqref{ns.e} since they are necessary and
sufficient -- it is simply much easier to obtain this result
directly from the Killing spinor equations). We also have
$\mathcal{L}_{\tilde{K}^2}\lambda=0$ (see~\eqref{warpok}) and
$\mathcal{L}_{\tilde{K}^2}\zeta= \mathcal{L}_{\tilde{K}^2}K^1= 0$
where $\mathcal{L}_{\tilde{K}^2}$ is the Lie derivative with
respect to $\tilde{K}^2$. This allows us to refine the local
coordinates $w^M=(x^i,\psi)$ with $i=1,\dots,4$, where as a vector
$\tilde{K}^2=3m\partial/\partial\psi$, and the factor of $3m$ has
been inserted for later convenience. Hence
\begin{equation}
   K^2 = \frac{1}{3m}\cos\zeta \left( \dd\psi + \rho \right)
\end{equation}
with $\rho=\rho_i(x^j,y)\dd x^i$. Locally the metric is then a product of a
four-dimensional space $M_4$ and the $K^1$ and $K^2$ directions:
\be
   \dd s^2  =  g^4_{ij} (x,y) \dd x^i \dd x^j
      + \me^{-6\lambda(x,y)}\sec^2\zeta(x,y) \dd y^2
      + \frac{1}{9m^2}\cos^2\zeta(x,y) (\dd\psi + \rho)^2
\label{itsgettinglate}
\ee
with $\lambda$ and $\zeta$ independent of $\psi$.
Note also, from~\p{ns.d},
that $i_{\tilde{K}^2}*G=-6\me^{3\lambda}\diff\lambda$ and hence
\be
   {\cal L}_{\tilde K^2} G =0~.
\ee
In other words, the Killing vector generates a symmetry not only
of the metric, and the $AdS_5$ warp factor $\lambda$, but also of
the four-form flux $G$.

Now let us turn to the metric $g^4$ on the four-dimensional part
of the space $M_4$. The forms $J=\frac{1}{2}J_{ij}\dd x^i\wedge\dd
x^j$ and $\Omega=\frac{1}{2}\Omega_{ij}\dd x^i\wedge\dd x^j$
define a local $SU(2)$ structure on $M_4$. Although the metric is
independent of $\psi$ it does not necessarily follow that $J$ and
$\Omega$ are. They also explicitly depend on $x^i$ and $y$. Let us
write
\begin{equation}
   \dd = \dd_4 + \dd y \wedge \partial_y
      + \dd\psi \wedge \partial_\psi~.
\end{equation}
It is also useful to define a rescaled structure and corresponding metric
\be
   J = \me^{-6\lambda}\hat J \qquad
   \Omega = \me^{-6\lambda}\hat \Omega \qquad
   g^4_{ij}=\me^{-6\lambda}\hat{g}_{ij}
\eea
so that the $d=6$ metric becomes
\be
   \dd s^2 =  \me^{-6\lambda(x,y)}\left[
         \hat{g}_{ij}(x,y) \dd x^i \dd x^j
         + \sec^2\zeta(x,y) \dd y^2 \right]
      + \frac{1}{9m^2}\cos^2\zeta(x,y) (d\psi + \rho)^2
\ee
Considering first $J$, the supersymmetry condition~\eqref{ns.d},
together with the four-form equation of motion~\p{eqmot}, imply
\begin{equation}
\label{jexpr}
\begin{aligned}
   \me^{-6\lambda}\diff (\me^{6\lambda}J) &=
         K^1 \wedge\left(
            \diff\log\cos\zeta \wedge K^2 -\diff K^2
            \right)\sin\zeta \\
         &=  -\frac{1}{3m}\diff \rho \wedge K^1 \cos\zeta \sin\zeta~.
\end{aligned}
\end{equation}
Decomposing, we find
\begin{align}
   \diff_4 \hat{J} & = 0 \label{Jint} \\
   \de_y \hat{J} & = -\frac{2}{3}y \, \diff_4\rho
      \label{defdrho} \\
   \de_\psi \hat{J} & = 0 ~.\label{killj}
\end{align}
For $\Omega$ we can work directly from the condition~\eqref{ns.c}. We
find
\begin{align}
   \diff_4 \hat{\Omega} & =
      (\ii \rho - \diff_4 \log\cos\zeta) \wedge \hat{\Omega}
      \label{barerho} \\
   \de_y \hat{\Omega} & =
      \left( - \tfrac{3}{2}y^{-1}\tan^2\zeta - \de_y\log\cos\zeta
         \right) \hat{\Omega} \label{deyOmega} \\
   \de_\psi \hat{\Omega} & =  \ii \hat{\Omega} ~.\label{killom}
\end{align}

We note in particular that $\dd\hat{\Omega}=A\wedge\hat{\Omega}$,
for a suitable one-form $A$. This implies that the almost complex
structure defined by $\hat{\Omega}$, or equivalently
$\hat{J}^i{}_j$, is independent of $\psi$ and $y$ and is
integrable on $M_4$. In other words, $M_4$ is a complex manifold.
Furthermore, from~\eqref{Jint}, we see that $\dd_4\hat{J}=0$, and
thus we have
\begin{equation}
   \text{$\hat{g}$ is locally a family of K\"ahler metrics on $M_4$
   parametrised by $y$.}
\end{equation}
The corresponding complex structure is independent of $y$ and
$\psi$, while from~\eqref{killj} we see the K\"ahler form $\hat{J}$
is independent of $\psi$.

Given the family of  K\"ahler metrics $\hat{g}$, we have the general
expression
\begin{equation}
   \dd_4 \hat{\Omega} = \ii \hat{P} \wedge \hat{\Omega}
\end{equation}
where $\hat{P}$ is the canonical Ricci-form connection defined by
the K\"ahler metric (at fixed $y$). That is, if the Ricci-form
$\hat{\Re}$ is defined by
$\hat{\Re}_{ij}=\frac{1}{2}\hat{R}^4_{ijkl}\hat{J}^{kl}$ then
\begin{equation}
   \hat{\Re} = \dd_4 \hat{P}~.
\end{equation}
Thus the content of
equation~\eqref{barerho} is to fix the one-form $\rho$ in terms of
$\hat{P}$ and $\dd_4\zeta$
\begin{equation}
\label{rho-rel}
 \rho = \hat{P} + \hat{J}\cdot \diff_4 \log\cos\zeta~.
\end{equation}
The $\de_\psi\hat{\Omega}$ condition is easily solved by redefining
the phase of $\hat{\Omega}$
\begin{equation}
\label{chargeom}
   \hat\Omega(x,y,\psi) = \me^{\ii \psi}\hat\Omega_0(x,y)~.
\end{equation}
The remaining content of the $\de_y\hat{\Omega}$ equation~\eqref{deyOmega}
is to fix the $y$-variation of the volume of $\hat{g}^4$. Recalling
that $\hat{\Omega}\wedge\bar{\hat{\Omega}}=4\widehat\vol_4$, we find
\begin{equation}
\label{vol-rel}
   \de_y\log\sqrt{\hat{g}} = -3y^{-1}\tan^2\zeta - 2\de_y\log\cos\zeta~.
\end{equation}
Note that compatibility of the last equation with
equation~\eqref{rho-rel} implies that\footnote{Recall that when
$\hat{g}$ is written in terms of four-dimensional complex
coordinates, defined by the $y$-independent complex structure
$\hat J_i{}^j$, locally we have $\hat{P}=\frac{i}{2}(\partial
-\bar\partial)\log\sqrt{\hat{g}}$.}
\begin{equation}
\label{1-form}
   \me^{3\lambda}\cos^3\zeta\de_y\rho = - 6m \hat{J}\cdot\dd_4\zeta
\end{equation}
or equivalently that
$\me^{3\lambda}\cos^3\zeta\de_y\rho-6m\ii\dd_4\zeta$ is a $(1,0)$-form
on $M_4$.

For a general two-form $\omega$ on $M_4$ we can define self-dual and
anti-self-dual combinations by
$\omega^{\pm}=\frac{1}{2}(\omega\pm\hat{*}_4\omega)$.
We have the identity
\be
   (\de_y \hat{J})^+ = \frac{1}{2}\de_y \log\sqrt{\hat{g}}\, \hat{J}
\ee
valid when the complex structure, $J_i{}^j$, is independent
of $y$. Given the relation~\eqref{defdrho}, we then see that the
condition on the volume~\eqref{vol-rel} can be written in the form
\bea\label{Jexplicit}
   (\diff_4\rho)^+ = 3m^2\me^{-6\lambda}\sec^2\zeta
      \left(1+6y\partial_y\lambda\right)\hat{J}~.
\eea

These conditions are in fact sufficient to ensure all the
relations \eqref{ns.a}--\eqref{ns.e} are satisfied. The only
remaining point is that one finds the flux $G$ is given by
\begin{equation}
\label{genflux}
\begin{aligned}
   G &= -(\partial_y \me^{-6\lambda})\widehat{\rm{vol}}_4
      - \me^{-9\lambda}\sec\zeta
         (\hat *_4\diff_4 \me^{6\lambda})\wedge K^1
      - \frac{1}{3m}\cos^3\zeta (\hat *_4\partial_y \rho)\wedge K^2 \\
      &\qquad
      + \me^{3\lambda}[\frac{1}{3m}\cos^2\zeta \hat*_4\diff_4\rho-4me^{-6\lambda}\hat J]
         \wedge K^1\wedge K^2~.
\end{aligned}
\end{equation}

In summary, we have shown that the necessary and sufficient
local conditions for $M_6$ to be supersymmetric and solve the equations
of motion are that the metric has the form
\begin{equation}
\label{g6}
   \dd s^2 =  \me^{-6\lambda}\left( \hat{g}_{ij} \dd x^i \dd x^j
         + \sec^2\zeta \dd y^2 \right)
      + \frac{1}{9m^2} \cos^2\zeta (d\psi + \rho)^2
\end{equation}
where $\hat{g}$, $\lambda$ and $\zeta$ and $\rho=\rho_i\dd x^i$
are all functions of $x^i$ and $y$. We also have
\begin{align}
   \text{(a)} & \quad\text{$\de/\de\psi$ is a Killing vector}
      \label{killingvec} \\
   \text{(b)} & \quad\text{$\hat{g}$ is a family of K\"ahler metrics
     on $M_4$ parameterized by $y$}
      \label{kahler} \\
   \text{(c)} & \quad\text{the corresponding complex structure
     $\hat{J}_i{}^j$ is independent of $y$ and $\psi$.}
      \label{integ}
\end{align}
In addition
\begin{align}
   \label{ydef}
   && && && \text{(d)} &&
      2m y &= \me^{3\lambda} \sin\zeta \\
   \label{rhodef}
   && && && \text{(e)} &&
      \rho &= \hat{P} + \hat{J}\cdot\dd_4\log\cos\zeta
   && && && &&
\intertext{where, in complex co-ordinates, $\hat{P}=\frac{1}{2}\hat{J}\cdot\dd\log\sqrt{\hat{g}}$ is the
Ricci-form connection defined by the K\"ahler metric, satisfying
$\hat{\Re}=\dd\hat{P}$. Finally we have the conditions}
   \label{dyJ}
   && && && \text{(f)}&&
      \de_y \hat{J} &= - \frac{2}{3} y \dd_4 \rho \\
   \label{dyg}
   && && && \text{(g)}&&
      \de_y\log\sqrt{\hat{g}} &=
         -3y^{-1}\tan^2\zeta - 2\de_y\log\cos\zeta
\end{align}
where, using~\eqref{dyJ}, the last expression can also be written as
\begin{equation}
\label{drho+}
   (\diff_4\rho)^+ = 3m^2\me^{-6\lambda}\sec^2\zeta
      \left(1+6y\partial_y\lambda\right)\hat{J}~.
\end{equation}
The four-form flux $G$ is given by~\eqref{genflux} and is
independent of $\psi$ -- that is, $\mathcal{L}_{\de/\de\psi}G=0$.
As shown in the previous subsection, the equations of motion for
$G$ and the Bianchi identity~\p{eqmot} are implied by
expressions~\eqref{killingvec}--\eqref{dyg}.

\section{Complex $M_6$ and explicit solutions}
\label{special}

In this section we consider how the conditions on the metric
specialise for solutions where the six-dimensional space $M_6$ is
a complex manifold. As we will see this additional condition is
equivalent to \be\label{allcondcom} \diff_4\zeta=0\qquad
\dd_4\lambda=0 \qquad \de_y\rho = 0 \ee and from the
condition~\eqref{rhodef}, we see that $\rho$ then coincides with
the canonical connection on the K\"ahler manifold $M_4$:
\begin{equation}\label{xfg}
    \rho = \hat{P}
 \end{equation}
Crucially, the supersymmetry conditions
simplify considerably and we are able to find many solutions in closed
form.  In particular, we find many new regular and compact
solutions that are topologically two-sphere bundles over the
K\"ahler base. These fall into two general classes and are
discussed in detail in sections~\ref{sec:KE}
and~\ref{sec:product}. In this section we first analyse what
the assumption that we have a complex structure implies about the
geometry of $M_6$ and then describe the global topology of the class
of solutions we consider.

\subsection{Conditions on local geometry}

We would like to specialize to the case where
\begin{equation*}
   \text{$\dd s^2(M_6)$ is a Hermitian metric on a complex
   manifold $M_6$.}
\end{equation*}
There is a natural almost complex structure compatible with $\dd
s^2(M_6)$ and the local $SU(2)$-structure given by complex
three-form $\Omega_{(3)}=\Omega\wedge(K^1+\ii K^2)$ (Note one
could equally well consider the three-form $\Omega\wedge(K^1-\ii
K^2)$ and get the same results). We find that $\dd\Omega_{(3)}$
has the form
\begin{equation}
\label{c-struct}
   \dd \Omega_{(3)} = A \wedge \Omega_{(3)}
      + v \wedge \Omega \wedge (K^1 - \ii K^2)
\end{equation}
where $v$ is a one-form given by
\begin{equation}
\label{vdef}
   v = - \left(\tan\zeta+\tfrac{1}{2}\cot\zeta\right)\dd_4\zeta
      + \ii\,\tfrac{1}{6m} \me^{3\lambda}\cos^2\zeta \de_y\rho~.
\end{equation}
In deriving~\eqref{c-struct} we have used the fact that
$\dd_4\rho\wedge\Omega=0$, which is a consequence
of~\eqref{rhodef}. For $\Omega_{(3)}$ to define an integrable complex
structure the second term in~\eqref{c-struct} must vanish. In general,
this implies $v\wedge\Omega=0$ or equivalently $v$ is a $(0,1)$-form
on $M_4$. This means that
\begin{equation}
   \me^{3\lambda}\cos^2\zeta \de_y\rho =
        - 6 m \left(\tan\zeta+\tfrac{1}{2}\cot\zeta\right)
           \hat{J}\cdot \dd_4\zeta~.
\end{equation}
However, we also have the integrability condition~\eqref{1-form}. The
only way both equations can hold is if $\dd_4\zeta = \de_y\rho = 0$
and, in fact, $v=0$. From~\eqref{ydef} we immediately have
$\dd_4\lambda=0$ and from~\eqref{rhodef} $\rho = \hat{P}$.
We thus get the results \p{allcondcom} and \p{xfg} stated above.

We will next derive conditions on the eigenvalues of the Ricci
tensor for the metric on the four-dimensional space $M_4$. By
definition, $\hat{\Re}=\dd_4\hat{P}$ so the conditions~\eqref{dyJ}
and~\eqref{drho+} now read
\begin{equation}
\label{Res}
\begin{aligned}
   \hat{\Re} & = - \frac{3}{2y}\de_y\hat{J} \\
   \hat{\Re}^+ &= 3m^2\me^{-6\lambda}\sec^2\zeta
      \left(1+6y\partial_y\lambda\right)\hat{J}~.
\end{aligned}
\end{equation}
Note that $\hat{\Re}^+$ is necessarily pointwise-proportional to
$\hat{J}$ since it is a self-dual  $(1,1)$-form. However, here we see
that the proportionality factor is independent of $x^i$. This implies
that the Ricci scalar
$\hat{R}\equiv \hat J^{ij}\hat\Re_{ij}=\hat J^{ij}\hat{\Re}^+_{ij}$
is constant:
\begin{equation}
\label{Rscalarconst}
   \diff_4 \hat R = 0~.
\end{equation}
Recall that on a K\"ahler manifold the Ricci tensor $\hat{R}_{ij}$
is related to the Ricci form by
$\hat{R}_{ij}=-\hat{J}_i{}^k\hat{\Re}_{kj}$. From the third
equation in~\eqref{allcondcom} and the fact that the complex
structure is independent of $y$, we have $\partial_y \hat
R_{ij}=0$. Note also that we can rewrite the first equation
of~\eqref{Res} as
\begin{equation}
   \hat{R}_{ij} =  -\frac{3}{2y}\de_y\hat{g}_{ij}~.
\end{equation}
Given $\de_y\hat{g}^{ij}=-\hat{g}^{ik}\hat{g}^{jl}\de_y\hat{g}_{kl}$
one finds
\begin{equation}
   \hat{R}_{ij}\hat{R}^{ij} = \frac{3}{2y}\de_y \hat{R}
\end{equation}
and hence
\begin{equation}
\label{normcurlyR}
   \dd_4 (\hat{R}_{ij}\hat{R}^{ij}) = 0~.
\end{equation}
For a K\"ahler metric the eigenvalues of
the Ricci tensor come in pairs, since it is invariant under the action
of the complex structure. Thus, in dimension four, there are \textit{a
  priori} two distinct eigenvalues which each have a degeneracy of
two. From~\eqref{Rscalarconst} we know that the sum of the
eigenvalues is a constant (on $M_4$). Moreover,
from~\eqref{normcurlyR} we see
that the sum of the squares of the eigenvalues is also constant. Thus
we find the useful condition that
\begin{equation*}
   \text{at fixed $y$, the Ricci tensor on $M_4$ has two pairs of
   constant eigenvalues.}
\end{equation*}
Finally, we note that the expression~\eqref{genflux} for the
flux $G$ simplifies to
\begin{equation}
   G = -\left(\de_y \me^{-6\lambda}\right) \widehat{\rm{vol}}_4
      + \me^{3\lambda}\left(
         \frac{1}{3m}\cos^2\zeta\, \hat{*}_4\diff_4 \rho
         - 4m \me^{-6\lambda}\hat J \right) \wedge K^1 \wedge K^2~.
\end{equation}

\subsection{Global structure of solutions}

We would like to find global regular solutions for the complex
manifold $M_6$. Our basic construction, which will cover almost
all regular compact solutions that we find, will be as follows. We
require that $\psi$ and $y$ describe a holomorphic $\bCP^1$ bundle
over a smooth K\"ahler base $M_4$
\begin{equation}
\label{fibration}
   \begin{CD}
      \bCP_{y,\psi}^1 @>>> M_6 \\
      && @VVV \\
      && M_4
   \end{CD}
\end{equation}
Recall that, given $\dd_4\zeta=0$, the metric has the form
\begin{equation}
\begin{aligned}
   \dd s^2 =&  \me^{-6\lambda(y)}\hat{g}_{ij}(x,y) \dd x^i \dd x^j \\
         &\qquad + \me^{-6\lambda(y)}\sec^2\zeta(y) \dd y^2
            + \frac{1}{9m^2}\cos^2\zeta(y)
               [\dd\psi + \hat{P}(x)]^2~.
\end{aligned}
\end{equation}
For the $(y,\psi)$ coordinates to describe a smooth $S^2$ we take
the Killing vector $\de/\de\psi$ to have compact orbits so that
$\psi$ defines an azimuthal angle. The coordinate $y$ is taken to
lie in the range $[y_1,y_2]$ with the $U(1)$ fibre, defined by
$\psi$, shrinking to zero size at the two poles $y=y_i$ -- that
is, we demand $\cos\zeta(y_i)=0$.

It is necessary to check that the fibre is smooth at the poles
$y=y_i$. As we shall see, in all cases in the above construction,
a smooth $S^2$ is obtained by choosing the period of $\psi$ to be
$2\pi$. (Note that this is consistent  with the integrated
expression~\eqref{chargeom} for $\Omega$). By definition,
$\hat{P}$ is a connection on the canonical bundle $\mathcal{L}$ of
the base $M_4$. Thus at fixed generic $y$, so that $y_1<y<y_2$,
the resulting 5-manifold is the total space of a $U(1)$ bundle
over $M_4$, which is in fact just the  
$U(1)$ bundle associated to the canonical line bundle
$\mathcal{L}$ of $M_4$. For the full $\bCP^1$ fibration we can
think of each fibre $\bCP^1$ as a projectivization of
$\bC^2=\bC\oplus\bC$. The transition functions of the $\bCP^1$
bundle act on the relative phase of the two factors of $\bC$. Thus
we can take one to be the trivial bundle $\mathcal{O}$. The other
is then $\mathcal{L}$. Projectivizing the bundles, we then have
that, as a complex manifold, $M_6$ is the total space of the
fibration
\begin{equation}
\label{topology}
   M_6 = \bP(\mathcal{O}\oplus\mathcal{L})~.
\end{equation}

We note that $M_6$ can also be viewed as the total space of the bundle
of unit self-dual\footnote{This is not to be confused with the
  bundle of unit \emph{anti}-self-dual two-forms, which is the twistor
  space of $M_4$.} two-forms over $M_4$. Here we think of each $S^2$
fibre as being a unit sphere in $\bR^3=\bR\oplus \bC$ with the
factor of $\bR$ being the polar direction on the $S^2$. Our $S^2$
bundle may then be viewed as the unit sphere bundle in an $\bR^3$
bundle, with the transition functions acting only in the $\bR^2 =
\bC$ part of the fibre. The rank 3 real bundle thus splits into a
direct sum $\mathcal{O}\oplus \mathcal{L}_{\bR}$ of a trivial real
line bundle $\mathcal{O}$, and (the realisation of) the complex canonical line
bundle $\mathcal{L}$. Consider now the two-forms on $M_4$. These
decompose into self-dual or anti-self-dual two-forms:
\bea
   \Lambda^2M_4 \cong \Lambda^+M_4\oplus \Lambda^-M_4~.
\eea
However, since $M_4$ is K\"ahler, the structure group of the tangent
bundle is in fact $U(2)\subset SO(4)$. Thus, one can now further
decompose the space of real two-forms as follows:
\bea
   \Lambda^+M_4 &\cong&
      \bR[\hat{J}]\oplus \mathcal{L}_{\bR}\nn
  \Lambda^-M_4 &\cong& \Lambda_0^{1,1}M_4~.
\eea
Here $\Lambda_0^{1,1}M_4$ denotes the bundle of primitive
$(1,1)$-forms {\it i.e.} 2-forms which are orthogonal to the K\"ahler
form $\hat{J}$, and are invariant under the action of the complex
structure.
Thus we see that the bundle of self-dual two-forms splits
as $\Lambda^+M_4 \cong \mathcal{O}\oplus \mathcal{L}_{\bR}$
where $\mathcal{O}$ is a trivial real line bundle generated by the
K\"ahler form on $M_4$. It is now clear that the $\bR^3$ bundle over
$M_4$ associated with our metrics is in fact the bundle of self-dual
two-forms. Notice that the ``polar direction'' is identified with the
direction corresponding to the K\"ahler form.

Let us now return to considering the base manifold $M_4$. We
showed in the previous subsection that, at fixed $y$, the Ricci
tensor on $M_4$ has two pairs of constant eigenvalues. We may now
invoke recent mathematical results from the literature on K\"ahler
manifolds. Theorem~2 of ref.~\cite{apostolov} states that, if the
Goldberg conjecture\footnote{The Goldberg conjecture says that any
compact Einstein almost K\"ahler manifold is K\"ahler-Einstein
i.e. the complex structure is integrable.} is true, then a compact
K\"ahler four-manifold whose Ricci tensor has two distinct pairs
of constant eigenvalues is locally the product of two Riemann
surfaces of (distinct) constant curvature. If the eigenvalues are
the same the manifold is by definition K\"ahler--Einstein. The
compactness in the theorem is essential, since there exist
non-compact counterexamples. However, for AdS/CFT purposes, we are
most interested in the compact case (for example, the central
charge of the dual CFT is inversely proportional to the volume).

Thus we naturally have two possibilities. First, the base $M_4$ is
K\"ahler--Einstein. The cosmological constant can depend on $y$,
so by definition we can write
\begin{equation}
   \text{case 1:} \qquad
   \hat{\Re} = \frac{k}{F(y)} \hat{J}
\end{equation}
where $k\in\{0,\pm1\}$ and $F(y)>0$.

Second, at fixed $y$, the base is a product of two constant curvature
Riemann surfaces. Thus the metric splits as
\begin{equation}
   \dd\hat{s}^2(M_4) = \dd\hat{s}^2_1 + \dd\hat{s}^2_2~.
\end{equation}
Let $\hat{J}_i$ be the corresponding K\"ahler forms. Again the
cosmological constant on each Riemann surface can depend on $y$. By
definition we can write
\begin{equation}
   \hat{\Re}_1 = \frac{k_1}{F_1(y)}\hat{J}_1 \qquad
   \hat{\Re}_2 = \frac{k_2}{F_2(y)}\hat{J}_2
\end{equation}
where $k_i\in\{0,\pm1\}$ and $F_i(y)>0$. Thus on $M_4$ we have
$\hat{J}=\hat{J}_1+\hat{J}_2$ and
\begin{equation}
   \text{case 2:} \qquad
   \hat{\Re} = \hat{\Re}_1 + \hat{\Re}_2 =
      \frac{k_1}{F_1(y)} \hat{J}_1
         + \frac{k_2}{F_2(y)} \hat{J}_2~.
\end{equation}
Clearly when $k_1=k_2$ and $F_1=F_2$ we get special
examples of case~1 above.

Since $\de_y\hat{P}=\de_y\rho=0$, we have $\de_y\hat{\Re}=0$. Thus
we immediately see that the rescaled K\"ahler forms given by
\begin{equation}
   \tilde{J} = \frac{1}{F}\hat{J} \qquad
   \tilde{J}_i = \frac{1}{F_i}\hat{J}_i
\end{equation}
are independent of $y$. The corresponding rescaled metrics
$\dd\tilde{s}^2=(1/F)\dd\hat{s}^2$ and
$\dd\tilde{s}^2_i=(1/F_i)\dd\hat{s}^2_i$, are independent of $y$
and have fixed cosmological constants $k$ and $k_i$, respectively
\begin{equation}
   \tilde{\Re} = k \tilde{J} \qquad
   \tilde{\Re}_i = k_i \tilde{J}_i~.
\end{equation}
For the Riemann surfaces of case~2 we write the metrics as
$\dd\tilde{s}^2(C_{k_i})$. Depending on the value of $k_i$, they
are the standard metrics on the torus $C_0\equiv T^2$, the sphere
$C_1\equiv S^2$ and hyperbolic space\footnote{In this last case,
hyperbolic space $H^2$ is
  of course non-compact. One might also consider compact $H^2/\Gamma$,
  where $\Gamma$ is a discrete group of isometries of $H^2$, but in
  general we expect these to break supersymmetry. Exceptions do
  exist such as, for example, the solution discussed in
  section~\ref{MN1sec}.} $C_{-1}\equiv H^2$.
From the first equation in~\eqref{Res} we can also solve for the
functions $F$ and $F_i$. We find
\begin{equation}
\begin{aligned}
   F(y) &= \tfrac{1}{3}(b-ky^2) \\
   F_i(y) &= \tfrac{1}{3}(a_i - k_iy^2)
\end{aligned}
\end{equation}
where $b$ and $a_i$ are constants. We then satisfy all the
constraints of supersymmetry except for the
$\de_y\log\sqrt{\hat{g}}$ condition~\eqref{dyg}.

Thus we have found that the metric on $M_4$ is given by
\begin{equation}
\begin{aligned}
   \text{case 1:} \qquad \dd s^2(M_4) &=
      \tfrac{1}{3}\me^{-6\lambda}\left(b-ky^2\right)\dd\tilde{s}^2(M_4)
      \\*[0.3cm]
   \text{case 2:} \qquad \dd s^2(M_4) &=
      \tfrac{1}{3}\me^{-6\lambda}\left(a_1-k_1y^2\right)
         \dd\tilde{s}^2(C_{k_1}) \\ & \qquad \qquad
      + \tfrac{1}{3}\me^{-6\lambda}\left(a_2-k_2y^2\right)
         \dd\tilde{s}^2(C_{k_2}) \\
\end{aligned}
\end{equation}
where $\me^{3\lambda(y)}\sin\zeta(y)=2my$ and the K\"ahler metrics
$\dd\tilde{s}^2(M_4)$ and $\dd\tilde{s}^2(C_{k})$ each satisfy
\begin{equation}
\label{KECOND}
   \tilde{\Re} = k \tilde{J}
\end{equation}
and are independent of $y$. The remaining equation~\eqref{dyg}, or
equivalently the second equation in~\eqref{Res}, implies
\begin{equation}
\label{diffeq}
\begin{aligned}
   \text{case 1:} &\qquad
   3m^2 F(1+6y\de_y\lambda) = k\, (\me^{6\lambda}-4m^2y^2) \\
   \text{case 2:} &\qquad
   6m^2 F_1F_2(1+6y\de_y\lambda) =
      (k_1F_2+k_2F_1)\,(\me^{6\lambda}-4m^2y^2)~.
\end{aligned}
\end{equation}

In summary, our basic construction is to solve these final equations
and look for solutions with a smooth base $M_4$ and a smooth $\bCP^1$
fibre with the topology given in equation~\eqref{topology}. We will
consider both compact and non-compact examples, although the theorem
leading to a consideration of these two classes of base manifold only
applies in the compact case. Note it is also possible to find smooth
metrics on $M_6$ when both the base and fibre metrics degenerate. One
example of this more general class is described in
section~\ref{sec:product}. In this case, the topology of the $U(1)$
fibration given by $\psi$ also changes.

\section{Case 1: K\"ahler-Einstein base}
\label{sec:KE}

We start by considering the case where the base is K\"ahler--Einstein
(KE). Recall that the metric has the form
\begin{equation}
\label{KEmetric}
   \dd s^2 =
      \tfrac{1}{3}\me^{-6\lambda}\left(b-ky^2\right)\dd\tilde{s}^2(M_4)
        + \me^{-6\lambda}\sec^2\zeta \dd y^2
            + \frac{1}{9m^2}\cos^2\zeta(\dd\psi + \tilde{P})^2
\end{equation}
where $\dd\tilde s^2(M_4)$ is a $y$-independent KE metric
on the four-dimensional base satisfying $\tilde{\Re} =
\dd_4 \tilde{P}=k\tilde{J}$. The remaining supersymmetry
condition~\eqref{diffeq} can be integrated explicitly. One finds, for
$k=\pm1$
\begin{equation}
\label{warpke}
\begin{aligned}
   \me^{6\lambda} &= \frac{2m^2(kb-y^2)^2}{cy+2kb+2y^2} \\
   \cos^2\zeta &= \frac{-3y^4-2cy^3-6kby^2+b^2}{(kb-y^2)^2}~.
\end{aligned}
\end{equation}
The solution is fully specified by giving the four-form flux which reads:
\begin{multline}
   G = \frac{(4y^3+3cy^2+12kby+kbc)}{18m^2(y^2-kb)} \,\tilde\vol_4 \\
        + \frac{k(y^4-6kby^2-2kbcy-3b^2)}{9m^2(y^2-kb)^2}\,
           \tilde{J}\wedge \diff y \wedge (\diff \psi + \tilde{P})~.
\end{multline}
For $k=0$, without loss of generality we can set $b=3$, and we
have simply
\begin{equation}
   \me^{6\lambda} = \frac{1}{cy} \qquad
   \cos^2\zeta = 1 - 4m^2cy^3~.
\end{equation}
The four-form flux for $k=0$ will be given shortly.

By construction these give locally supersymmetric solutions. We
next investigate the solutions in more detail, in particular,
analysing when they are regular. We will find that there are
regular solutions only for $k=1$. An interesting
aspect of the $k=0$ case, when the hyper-K\"ahler manifold is
taken to be a four-torus, is that the solutions can be reduced to
type IIA solutions and thence, on performing a T-duality, to
(singular) Sasaki-Einstein type IIB solutions.

\subsection{KE bases with positive curvature: $k=1$}
\label{KEk=1}

We consider the case where $k=1$. From the form of the
metric~\eqref{KEmetric}, we see that we require $b>0$. Without
loss of generality, by an appropriate rescaling of $y$ we can set
$b=1$. In addition, by flipping the sign of $y$ if necessary, we
can take $c$ to be positive. The warp factor $\me^{6\lambda}$
becomes zero at $y^2=1$ and the metric on $M_4$ also develops a
singularity, and thus we require
\begin{equation}
   y^2 < 1~.
\end{equation}

First we consider $0\le c < 4$, in which case $\cos^2\zeta$
has two zeroes, at $y_1, y_2$, which are the two real roots of the
quartic:
\be\label{quartic}
   3y^4+2cy^3 +6y^2-1=0~.
\ee
We take $y_1\le y\le y_2$ and note the remarkable fact that within
this range, the  expression for $\cos^2\zeta$ is consistent with
$\zeta$ running monotonically from $-\pi/2$ to $\pi/2$ (for
example $\cos^2\zeta\le 1$ and is equal to 1 just at $y=0$). We
also find that $y^2<1$ and so the base metric is well-defined.

Now, if $\psi$ is periodic, the $(y,\psi)$ part of the metric is a
metric on a non-round two-sphere provided that a single choice for
the period of $\psi$ ensures the metric is smooth at both of the
poles, located at $y=y_1,y_2$.  Let $\Sigma_{1}$ be the value of
the derivative of  $\cos^2\zeta$ at $y_{1}$. Assuming that
$\Sigma_1$ is non-vanishing, which will be the typical case, then
near $y_1$ the $(y,\psi)$ part of the metric becomes
\be
   \frac{\me^{-6\lambda(y_1)}}{\Sigma_1(y-y_1)}\dd y^2
     + \frac{1}{9m^2}\Sigma_1(y-y_1)(\dd\psi+\tilde{P})^2~.
\ee
If we now introduce the co-ordinate $\epsilon=2(y-y_1)^{1/2}$,
this can be written
\be
   \frac{\me^{-6\lambda(y_1)}}{\Sigma_1}\left[\dd\epsilon^2
      + \alpha^2\epsilon^2(\dd\psi+\tilde{P})^2
         \right]
\ee
where
\begin{equation}
   \alpha^2 = \frac{1}{36m^2}\me^{6\lambda(y_1)}\Sigma_1^2~.
\end{equation}
This is regular provided that $\psi$ has period $2\pi\alpha$.
Remarkably, for \emph{both} $y_1$ and $y_2$ (in fact for all the
roots of the quartic~\p{quartic}) we find $\alpha^2=1$. Thus
choosing the period of $\psi$ to be $2\pi$ implies that we do
indeed have a smooth two-sphere. This is compatible with the fact
that $\tilde{P}$ is the K\"ahler connection on the canonical
bundle $\mathcal{L}$, and implies that at fixed $y\neq y_i$ the
five-dimensional manifold is the total space of the associated
$U(1)=S^1$ bundle to $\mathcal{L}$. In addition, it is compatible
with the solution~\p{chargeom} for the two-form $\Omega$, implying
that $\Omega$ has charge one under the action of the Killing
vector $\de/\de\psi$. In other words, for $0\le c<4$ we have a
one-parameter family of completely regular solutions which are
$S^2$ fibrations over a smooth KE base.

Next consider $c=4$. This case needs a special analysis since the
quartic develops a triple root at $y=-1$. Another root is located
at $y=1/3$. As before, choosing the period of $\psi$ to be $2\pi$
leads to a smooth metric near to $y=1/3$. On the other hand, this
period leads to a conical singularity at $y=-1$. This is in fact a
curvature singularity in general, the exception being when the
base is $\mathbb{C}P^2$, in which case one has an orbifold
singularity at this point. In fact the solution is in this case
the weighted projective space $W\mathbb{C}P^3_{[1,1,1,3]}$.

Finally, when $c>4$, the quartic only has a single real positive
root. Once again, choosing the period of $\psi$ to be $2\pi$ leads
to a smooth metric at $y$ equal to this root. On the other hand,
the space includes the point $y=-1$ where the warp factor
$\me^{6\lambda}$ is zero and the metric is singular.

In summary, we have found that:
\begin{quote}
\begin{itshape}
   For $0\leq c < 4$ we have a one-parameter family of completely
   regular, compact, complex solutions with the topology of a $\bCP^1$
   fibration over a positive curvature KE space.
\end{itshape}
\end{quote}
Since four-dimensional compact K\"ahler-Einstein spaces with
positive curvature have been classified~\cite{tian,tianyau}, we
have a classification for the above solutions. In particular, the
base space is either $S^2\times S^2$ or $\mathbb{C}P^2$, or
$\mathbb{C}P^2\#_n \mathbb{C}P^2$ with $n=3,\dots,8$. For the
first two examples, the KE metrics are of course explicitly known
and this gives explicit solutions of M-theory when fed into the
above solutions. The remaining metrics, although proven to exist,
are not explicitly known, and so the same applies to the
corresponding M-theory solutions.

\subsection{KE bases with negative curvature: $k=-1$}

We now consider cases with $k=-1$. First consider the case
$b\neq0$. Arguing as above, by redefining $y$ we can set $b=1$ and
we may also take $c$ to be positive without loss of generality. To
ensure that $\me^{6\lambda}$ is positive we need to take
\begin{equation}
   y\ge y_+ \equiv \frac{1}{4}\left(-c+\sqrt{c^2+16}\right),\quad
\text{or} \quad  y\le y_- \equiv
\frac{1}{4}\left(-c-\sqrt{c^2+16}\right).
\end{equation}
Now, for all values of $c$ we find that $\cos^2\zeta$ has two real
roots, $y_1$ and $y_2$. The largest one, $y_2$, is always greater
than $y_+$, while $y_1$ is always smaller $y_-$. To ensure that
$\cos^2\zeta$ is positive, we must take the range of $y$ to be
either: $y_+\le y \le y_2$ or $y_1\le y \le y_-$. When
$y=y_{1,2}$, in each of the two solutions, by carrying out an
analysis as above we find that the metric is regular provided that
the period of $\psi$ is once again taken to be $2\pi$. However,
when $y=y_\pm$, the warp factor $\me^{6\lambda}$ goes to infinity
and we have a singularity.

When $b=0$ we can again take $c\ge 0$, without loss of generality.
For $\me^{6\lambda}$ to be positive and finite, we require
$y<-\frac{1}{2}c$ or $y>0$. The roots of $\cos^2\zeta$ are now
$y=0$ and $y=-\frac{2}{3}c$. Thus the possible ranges of $y$ are
either $y\geq 0$, in which case the solution is singular at $y=0$,
or $-\frac{2}{3}c\le y<-\frac{1}{2}c$ with $c$ strictly positive,
$c>0$. In the latter case, while $y=-\frac{2}{3}c$ is regular if the
period of $\psi$ is taken to be $2\pi$, at $y=-\frac{1}{2}c$ again the
solution is singular.

\subsection{Hyper-K\"ahler bases: $k=0$}

Now we consider the case $k=0$. This means the base is a
Ricci-flat K\"ahler manifold, or in other words the
$y$-independent four-dimensional metric $\tilde g$ is
hyper-K\"ahler. Locally we can, and will, choose $\tilde P=0$. As
discussed above the solution has warp factor $\me^{6\lambda}=1/cy$
and hence $\cos^2\zeta=1-4m^2cy^3$. Without loss of generality we
can assume $c\geq0$ and hence $y>0$ for the warp factor to be
positive.

After rescaling $y\rightarrow y/c^{1/3}$,  and $\tilde{g}\to
c^{-2/3}\tilde{g}$, the full eleven-metric takes the simple form
\begin{multline}
\label{K3}
   \dd s^{2}_{11} = c^{-2/9}\bigg[
      \frac{1}{y^{1/3}} \dd s^{2}(AdS_{5})
      + y^{2/3}\dd \tilde{s}^{2}(M_4) \\
      + \frac{y^{2/3}}{1-4{m^2}y^{3}} \dd y^{2}
      + \frac{1}{9m^2y^{1/3}}\left(1-4m^2y^{3}\right)
         \dd\psi^{2}
      \bigg]
\end{multline}
where, if the hyper-K\"ahler four-metric $\dd\tilde{s}^2$ is to be
compact, the base $M_4$ is either $T^{4}$ or $K3$ (modulo a quotient
by a freely acting finite group). The flux takes the form
\begin{equation}
   G = c^{-1/3}\left[
      -\widetilde{\mathrm{vol}}_{4}
      - \frac{4}{3}y \,\diff y \wedge \diff \psi \wedge \tilde{J} \right]
\end{equation}
where $\tilde{J}$ and $\widetilde{\vol}_{4}$
denote the K\"ahler form and volume form on the hyper-K\"ahler space,
respectively.  We clearly see that the constant $c$ is just an overall
scaling and is not a genuine modulus, and so henceforth we set $c=1$.

If we introduce the new variable:
\begin{equation}
\begin{aligned}
   y &= \frac{1}{(4m^2)^{1/3}}\cos^{2/3}\theta
\end{aligned}
\end{equation}
the $D=11$ solution becomes
\begin{equation}
\begin{aligned}
   \dd s^{2}_{11} &= y^{-1/3}\left[
      \dd s^{2}(AdS_{5})
      + \frac{1}{9m^2}\left(
         \dd\theta^2 +\sin^2\theta \dd\psi^2\right)
         \right]
      + y^{2/3} \dd\tilde{s}^{2}(M_4) \\
   G &= -\widetilde{\mathrm{vol}}_{4}
      + \frac{4}{9m}y^{1/2}\sin\theta \,\diff\theta
      \wedge \diff \psi \wedge \tilde{J}~.
\end{aligned}
\end{equation}
If the period of $\psi$ is $2\pi$, then the solution is regular at
$\theta=0$. On the other hand, $\theta=\pi/2$ is clearly singular.

In the special case that the HK space is flat, we can obtain
supersymmetric solutions of type IIA and type IIB supergravity by
dimensional reduction and $T$-duality, respectively. Note that
these solutions will be supersymmetric provided that the Killing
spinor is invariant under the action of the Killing vector that one is
reducing on or T-dualising on. This will be the case for the flat
torus directions but will not in general be the case if we reduce on
$\psi$.  If we write the flat HK metric as
$\dd x_1^2+\dd x_2^2+\dd x_3^2+\dd x_4^2$ with $\tilde J= \dd x_1\wedge \dd x_2 + \dd x_3\wedge \dd x_4$ and then dimensionally reduce on the $x_4$ direction we obtain the type IIA supersymmetric solution
\begin{equation}
\begin{aligned}
   \dd s^2 &= \dd s^2(AdS_5)
      + \frac{1}{9m^2}\left(
         \dd\theta^2 +\sin^2\theta\dd\psi^2\right)
      + y\left(\dd x_1^2+\dd x_2^2+\dd x_3^3\right) \\
   \me^{2\Phi} &= y \\
   F^{RR} &= \frac{4}{9m}y^{1/2}\sin\theta\,
      \diff\theta\wedge\diff{\psi}\wedge\dd x_1\wedge\dd x_2 \\
   B^{NS} &= -\frac{1}{2}x_1 \diff x_2\wedge \diff x_3
      + \frac{1}{2}x_2 \diff x_1\wedge \diff x_3
      - \frac{2}{3}y^2 \diff \psi\wedge \diff x_3
\end{aligned}
\end{equation}
where $\dd s^2$ is the type IIA string metric, $\Phi$ is the
dilaton, $F^{RR}$ is the Ramond--Ramond four-form field strength
and $B^{NS}$ is the Neveu--Schwarz two-form potential. Here we are
using the conventions of, for example, ref.~\cite{hkpap}.

If we T-dualise now on the $x^3$ direction, using the formulae in
ref.~\cite{Bergshoeff:1995as}, we obtain a supersymmetric type IIB
solution whose metric is the direct product of $AdS_5$ with a
five-manifold $X_5$. The solution has constant dilaton and the
only other non-vanishing field is the self-dual five-form. The
metric on $X_5$ is given by
\begin{equation}
\begin{aligned}
   \dd s^2(X_5) &= y\left( \dd x_1^2+ \dd x_2 ^2\right)
      + \frac{\sin^2\theta}{4y} \left(
         2\dd x_3+x_2\dd x_1-x_1 \dd x_2\right)^2 \\ &\qquad
      + \frac{1}{9m^2}\dd\theta^2
      + \frac{1}{9m^2}\left[
         \dd\psi - 3m^2y(2\dd x_3+x_2\dd x_1-x_1\dd x_2)
      \right]^2~.
\end{aligned}
\end{equation}
For this to be a supersymmetric solution of type IIB string theory
we know that the metric should be Sasaki-Einstein. It is
satisfying to check that the Ricci tensor is given by $4m^2$ times
the metric. A calculation of the square of the Riemann tensor is
given by $8m^4(5+48\cos^{-4}\theta)$, which implies that the
solution is singular at $\theta=\pi/2$.

\section{Case 2: Product base}
\label{sec:product}

Now we consider the case where the base is a product of two
constant curvature Riemann surfaces. Recall that the metric has
the form
\begin{multline}
\label{prod-metric}
   \dd s^2 =
      \tfrac{1}{3}\me^{-6\lambda}
         \left(a_1-k_1y^2\right)\dd\tilde{s}^2(C_1)
      + \tfrac{1}{3}\me^{-6\lambda}
         \left(a_2-k_2y^2\right)\dd\tilde{s}^2(C_2) \\
      + \me^{-6\lambda}\sec^2\zeta \dd y^2
            + \frac{1}{9m^2}\cos^2\zeta\left(d\psi + \tilde{P}\right)^2
\end{multline}
where the $y$-independent metrics $\dd\tilde{s}^2(C_i)$ describe 
constant curvature Riemann surfaces with curvature
$k_i\in\{0,\pm1\}$. In other words the metrics on $C_i$ are the
standard ones on either tori $T^2$, spheres $S^2$ or hyperbolic
spaces $H^2$.

We must solve the remaining supersymmetry condition~\eqref{diffeq}
to find $\lambda$ and $\zeta$ as functions of $y$. Substituting
for $F_i$, the condition can be written as
\begin{multline}
\label{easytosolve}
   2m^2y\partial_y \me^{6\lambda}\left(a_1-k_1y^2\right)
         \left(a_2-k_2y^2\right)
      = \me^{12\lambda}\left(k_1a_2 + k_2 a_1 - 2k_1k_2y^2\right) \\
         + 2m^2\me^{6\lambda}\left(
            3k_1k_2y^4 - (k_1a_2+k_2a_1)y^2 -a_1a_2\right)~.
\end{multline}
Remarkably, this can again be solved in closed form for general
$k_i$. In the following we will treat two cases separately. First
we note that $k_1=k_2=0$ is just a four-torus and so this case was
considered in the previous section. We can then distinguish
between the case where one of the $k_i$ is zero and where neither
is zero. We first consider the former case, when one of the
Riemann surfaces is a flat two-torus. For this case we will also
show that, by dimensionally reducing on one of the circle
directions, we can obtain additional type IIA solutions, and that
a further T-duality on the other circle direction leads to type
IIB solutions. In particular, in this way we obtain new
Sasaki-Einstein spaces.

\subsection{$S^2\times T^2$ and $H^2\times T^2$ base: $k_2=0$}
\label{tdualsols}

Let us first suppose that one of the Riemann surfaces is flat. Then
without loss of generality we may set $k_1=k=\pm 1$, $k_2=0$ and also write
 $a_1=a$.  The general solution to~\eqref{easytosolve} is then:
\begin{equation}
\begin{aligned}
   \me^{6\lambda} &= \frac{2m^2(a-ky^2)}{k-cy} \\
   \cos^2\zeta &= \frac{a-3ky^2+2cy^3}{a-ky^2}
\end{aligned}
\end{equation}
where $c$ is an integration constant.
Without loss of generality we can set $a_2=3$.
The full supersymmetric solution can then be written
\begin{equation}
   \dd s^2 = \me^{-6\lambda}\dd\tilde{s}^2(T^2)
      + \frac{k-cy}{6m^2}\dd\tilde{s}^2(C_k)
      + \me^{-6\lambda}\sec^2\zeta\dd y^2
      + \frac{1}{9m^2}\cos^2\zeta(\dd\psi+\tilde{P})^2
   \label{s2s2t2}
\end{equation}
where $\tilde{P}$ denotes the canonical connection for the metric
$\dd\tilde{s}^2(C_k)$. The constants $a$ and $c$ are arbitrary. We
find that the flux is given by
\begin{multline}
   G = \frac{-2y+ky^2c+ca}{6m^2(a-ky^2)}
         \vol(C_k)\wedge\vol(T^2)
      - \frac{2(k-cy)}{9m^2}
         \diff y \wedge(\diff \psi+\tilde{P})\wedge\vol(C_k) \\
      - \frac{ak+y^2-2acy}{3m^2(a-ky^2)^2}
         \diff y\wedge (\diff \psi+\tilde{P})\wedge\vol(T^2)
\end{multline}
where $\vol(C_k)$ is the standard volume form on $C_k\in\{S^2,H^2\}$
for $k=\pm 1$.

We are looking for solutions which are $S^2$ fibrations over
$C_k\times T^2$. For compact fibres, as before, we consider
$y_1\leq y\leq y_2$ where $y_i$ are two roots of $\cos^2\zeta=0$
giving the poles of the $S^2$. Moreover, we also require that
$k-cy$ and $\me^{-6\lambda}$ should both be positive in this range
so that the metric on $C_k\times T^2$ does not have singularities.
Finally we must check that the poles of the fibre $S^2$ are free
of conical singularities for a suitable choice of the period of
$\psi$.

To investigate whether such solutions exist, first consider $c\ne 0$.
In this case we can set $c=1$ without loss of generality.  For $k=1$, we
have
\begin{equation}
\begin{aligned}
   \dd\tilde{s}^2(S^2) &= \dd\theta^2 + \sin^2\theta \dd\phi^2\nn
   \tilde{P} &= - \cos\theta \dd\phi ~.
\end{aligned}
\end{equation}
In this case, for the range $0<a<1$ the cubic in the numerator of
$\cos^2\zeta$ has three real roots. If we choose $y$ to lie within
the range between the two smallest roots we find that all of the
conditions mentioned above are satisfied, if we choose the period
of $\psi$ to be $2\pi$. In other words, for $0<a<1$ we have a
two-parameter family of completely regular, compact, complex
solutions that are $S^2$ bundles over $S^2\times T^2$. For $k=-1$
we find that the analogous solutions with $-1<a<0$ are singular.

For $k=1$, still with $c=1$, if one sets $a=0$ it is
straightforward to see that the resulting space has curvature
singularities. The solution with $a=1$ is also singular, but in a
milder and more interesting way. It is convenient to perform a
change of variables, defining
\begin{equation}
   y =  1 -\frac{3}{2}\sin^2 \sigma
\end{equation}
so that the $a=1$ metric becomes
\begin{multline}
   m^2 \dd s^2 = \dd\sigma^2
      + \frac{1}{4}\sin^2\sigma (\dd\theta^2+\sin^2\theta\dd\phi^2) \\
      + \frac{\sin^2\sigma\cos^2\sigma}{1+3\cos^2\sigma}
         (\dd\psi - \cos\theta d\phi)^2
      + \frac{1}{1+3\cos^2\sigma}\dd\tilde{s}^2(T^2)~.
\end{multline}
The range of $\sigma$ is $0\leq\sigma\leq\pi/2$. It is easy to see
that near $\sigma =\frac{1}{2}\pi$ one smoothly approaches a bolt
$T^2\times S^2$ of co-dimension two if and only if ${\psi}$ has
period $2\pi$. On the other hand, near to $\sigma=0$ the metric
collapses to a co-dimension four bolt $T^2$. At constant $\sigma$,
the collapsing fibre turns out to be $\mathbb{R}P^3=S^3/Z_2$,
rather than $S^3$, due to the periodicity already enforced on
$\psi$. In fact, projecting out the $T^2$, the resulting
four-space is the weighted projective space
$W\mathbb{C}P^2_{[1,1,2]}$. This is a complex orbifold. Notice the
close similarity of the metric we have to the standard
K\"ahler--Einstein metric on $\mathbb{C}P^2$.

If one now starts with $W\mathbb{C}P^2_{[1,1,2]}$, one can
consider blowing up the $\mathbb{Z}_2$ singularity, replacing it
locally with $T^*S^2$. The resulting space is then clearly two
copies of the cotangent bundle of $S^2$ glued back to back. This
is precisely the topology of the non-singular spaces described
above, when $0<a<1$. In fact, one can show that the resulting
$S^2$ bundle over $S^2$ is topologically trivial. Thus we can
think of the limit $a\rightarrow 1$ as a limit in which we blow
down an $S^2$ in the non-singular family to obtain an orbifold,
and it follows that the parameter $a$ measures the size of this
cycle, with $a\rightarrow 1$ the zero-size limit.

The remaining case to consider is $c=0$. Clearly for the metric to
have the correct signature we must have $k=1$. Without loss of
generality, we can also set $a=3$. With the change of coordinates
$y=\cos\omega$ we find
\begin{multline}
   6m^2 \dd s^2 = \dd\theta^2
      + \sin^2\theta\dd\phi^2 + \dd\omega^2 \\
      + \frac{2\sin^2\omega}{2+ \sin^2\omega}
         \left(\dd{\psi}-\cos\theta\dd\phi\right)^2
      + \frac{3}{2+\sin^2\omega}\dd\tilde{s}^2(T^2)~.
\end{multline}
where $0\leq\omega\leq \pi$. The total space is a regular $S^2$
bundle over $S^2\times T^2$ provided that the period of $\psi$ is
taken to be $2\pi$. In this case, projecting out the $T^2$, the
resulting space is an $S^2$ bundle over $S^2$, which furthermore
is topologically trivial (for more details, see the discussion in
ref.~\cite{paper2}).

In summary, we see that given $k=1$ we have:
\begin{quote}
\begin{itshape}
   For $0<a<1$ and $c\neq 0$ we have a one-parameter family
   of completely regular, compact, complex solutions that are
   topologically trivial $S^2$ bundles over $S^2\times T^2$. A
   single additional solution of this type is obtained when $c=0$ and
   $a\neq 0$.
\end{itshape}
\end{quote}
There are no non-singular solutions for $k=-1$.

\subsection{New type IIA and type IIB solutions}

This family of solutions also leads to solutions of type IIA and
IIB supergravity. First, for any allowed value of the constants
$a$ and $c$, we can reduce the above solutions on one of the torus
directions to obtain supersymmetric type IIA solutions.
Concretely, writing $\dd\tilde{s}^2(T^2)=\dd x_3^2+\dd x_4^2$ and
reducing along the $x_4$-direction we get
\begin{equation}
\begin{aligned}
   \dd s^2 &= \dd s^2(AdS_5)
      + \frac{k-cy}{6m^2}\dd\tilde{s}^2(C_k)
      + \me^{-6\lambda}\sec^2\zeta\dd y^2
      \\ & \qquad \qquad \qquad \qquad \qquad
      + \frac{1}{9m^2}\cos^2\zeta(\dd\psi+\tilde{P})^2
      + \me^{-6\lambda}\dd x_3^2 \\
   \me^{2\phi} &= \me^{-6\lambda} \\
   G^{RR} &= -\frac{2(k-cy)}{9m^2}\diff y\wedge
      (\diff \psi + \tilde{P})\wedge\vol(C_k) \\
   B^{NS} &= \frac{1}{6m^2k}\frac{-2y+ky^2c+ca}{a-ky^2}
      (\diff \psi+\tilde{P})\wedge \diff x_3~.\label{iiasolu}
\end{aligned}
\end{equation}
For $k=1$, the $S^2$ bundle is still smooth and in fact
we have completely regular type IIA solutions for the values of $a$,
$c$ and the ranges of $y$ and $\psi$ where the $D=11$ solutions are
regular.

We can now T-dualise along the $x_4$-direction to get type IIB
solutions. The metric is given by
\begin{multline}
\label{tinky}
   \dd s^2 = \dd s^2(AdS_5)
      + \frac{k-cy}{6m^2}\dd\tilde{s}^2(C_k)
      + \me^{-6\lambda}\sec^2\zeta \dd y^2
      + \frac{1}{9m^2}\cos^2\zeta(d\psi+\tilde{P})^2 \\
      + {\me^{6\lambda}}\left[\dd x_3
         + \frac{1}{6m^2k}\frac{-2y+ky^2c+ac}{a-ky^2}
            (\dd\psi+\tilde{P})\right]^2~.
\end{multline}
We find that the dilaton is constant and the only other
non-vanishing field is the self-dual 5-form. Thus the solutions should
be the product of $AdS_5$ with a locally Sasaki--Einstein space. We
have checked that it is Einstein with the appropriate factor of
$4m^2$, and also locally Sasaki.

The global structure of these solutions for $k=1$ will be
discussed in detail in ref.~\cite{paper2}.  In particular, we find
the regular M-theory solution with $c=0$ leads to a regular type
IIA solution and the resulting Sasaki-Einstein metric arising in
the type IIB solution is simply the standard one on
$T^{1,1}/\mathbb{Z}_2$. The four-dimensional superconformal field
theory dual to this type IIB solution was identified in
ref.~\cite{Morrison:1998cs}.
In ref.~\cite{paper2} we will also show that for the
solutions with $0<a<1$ and $c\neq 0$, there are a countably infinite
number of values of $a$ for which the type IIB solution gives a
regular Sasaki-Einstein metric on $S^2\times S^3$.
These should correspond to new $AdS_5$ duals of $N=1$ superconformal field
theories and it will be interesting to see when the type IIB,
type IIA or M-theory supergravity solutions are most useful.
Interestingly, the mildly singular $D=11$ limiting solution  with
$a=1$ (and $c=1$) is dual to $S^5/\bZ_2$ in the type IIB theory.

\subsection{$S^2\times S^2$, $S^2\times H^2$ and $H^2\times H^2$
  bases: $k_i=\pm1$}

We now consider $k_i=\pm1$, corresponding to warped products of
$S^2 \times S^2$, $H^2\times H^2$, or $S^2 \times H^2$. Notice
that we should at least recover the KE result for the first two of
these cases, when the warping is the same for each space. In fact,
the warping can be different, corresponding to choosing the
integration constants $a_1\neq a_2$. The general solution
to~\eqref{easytosolve} is given by
\bea
   \me^{6\lambda} &=&
      \frac{2m^2(y^2-a_1k_1)(y^2-a_2k_2)}{2y^2+cy+k_1a_1+k_2a_2}\nn
   \cos^2 \zeta &=&
      \frac{-3y^4 -2cy^3-3(k_1a_1+k_2a_2)y^2+k_1k_2a_1a_2}
      {(y^2-k_1a_1)(y^2-k_2a_2)}
\eea
where $c$ is an arbitrary integration constant.
The four-form flux is given by
\bea
G &=& \frac{k_1k_2}{18m^2(y^2-a_1k_1)(y^2-a_2k_2)}\big[ 4y^5 +3cy^4
+4y^3(a_1k_1+a_2k_2)\nn &&-cy^2(a_1k_1+a_2k_2)
 -2 y(a_1^2+a_2^2+4a_1a_2k_1k_2)- ca_1a_2k_1k_2 \big]\vol_1\wedge \mathrm{vol}_2\
\nn
&& +\Big\{ \frac{k_2}{9m^2(y^2-a_1k_1)^2} \big[y^4-y^2(a_2k_2+5a_1k_1)-2a_1k_1cy\nn
&& -a_1a_2k_1k_2-2a_1^2\big]\mathrm{vol}_2 +\big[1 \leftrightarrow 2 \
\big]\mathrm{vol}_1\Big\}\wedge \diff y \wedge (\diff \psi + \tilde P)~.
\eea

We therefore have a three-parameter family of solutions. These
reduce, on setting $a_1=a_2=\pm b$, to the KE solutions considered
in section~\ref{KEk=1} with base $S^2\times S^2$ or $H^2\times
H^2$ (for the cases $k_1=k_2=k=\pm 1$, respectively).

As usual we are particularly interested in regular solutions where
the $(y,\psi)$ part of the metric is a two-sphere. This requires
that $y$ is bounded between two suitable roots $y_1$ and $y_2$ of
the quartic appearing in the numerator of $\cos^2\zeta$. As
before, we find the two-sphere will be free of conical
singularities if the period of $\psi$ is taken to be $2\pi$. A
regular solution is obtained if, for values of $y$ between the
roots, the warp factor $\me^{6\lambda}$ and the $F_i$ are positive
and $0\le\cos^2\zeta\le 1$. We find:

\begin{quote}
\begin{itshape}
 For various ranges of $(a_1,a_2,c)$ there are regular solutions that are
 topologically $S^2$ bundles over $S^2\times S^2$ and $S^2\times H^2$.
\end{itshape}
\end{quote}
In particular, for the $S^2\times S^2$ case there are solutions when
$a_1$ is not equal to $a_2$ and hence this gives a broader class of
solutions than in the K\"ahler-Einstein case considered above.
The existence of regular solutions is rather easy to see if one sets
$c=0$. It is also not difficult to show that when $c=0$ there are no
regular solutions of the type being considered when the base is
$H^2\times H^2$, since the positivity conditions cannot be satisfied.

For the $S^2\times H^2$ case there is a special degenerate limit
which leads to a solution first found in ref.~\cite{malnun}, as we
show in the next subsection. This solution is of a different
global type from those considered thus far. First, the $S^2$
fibres are not smooth but have conical singularities at the poles.
This is connected to the fact that the $U(1)$ bundle specified by
$\psi$ is globally $\mathcal{L}^{1/2}$ rather than $\mathcal{L}$.
In addition, the volume of the $S^2$ of the $S^2\times H^2$ base
goes to zero at the poles of the $(y,\psi)$ two-sphere. Together
these facts conspire to make the total space of $S^2$ and
$(y,\psi)$ a four-sphere. In other words, the topology of these
solutions is actually an $S^4$ bundle over $H^2$.
Note that, for the class of regular solutions described in the last
paragraph, the volume of the $S^2$ in the $S^2\times H^2$ base is
always finite and topologically we have an $S^2$ bundle over
$S^2\times H^2$. This combined with the fact that the ranges for the
co-ordinate $\psi$ are different indicate that they should probably
not be viewed as deformations of the solution found in
ref.~\cite{malnun}. Finally, we comment that in the solution of
ref.~\cite{malnun} it was argued that the Killing spinors do not
depend on the co-ordinates on $H^2$ and hence this space can be
replaced with $H^2/\Gamma$ where $\Gamma$ is a discrete group of
isometries. It would be interesting to know whether this is also
possible for our more general solutions with base $S^2\times
H^2$. This could be established with an explicit expression for the
Killing spinor.


\subsection{Recovering the $\mathcal{N}=1$ Maldacena-N\'u\~nez solution}
\label{MN1sec}

Given our general formulation, it is a simple exercise to recover
the regular solution with $\mathcal{N}=1$ supersymmetry
constructed by Maldacena and N\'u\~nez in ref.~\cite{malnun}. This
solution corresponds to setting $k_1=-k_2=1$, so that the base
four-manifold is a warped product of $S^2\times H^2$. Crucially
though, the $(y,\psi)$-fibre is \emph{not} a smooth $S^2$ and, in
addition, the base metric is singular at certain values of $y$.
Thus the topology of this solution is different to that of the
regular solutions we have considered so far.

For simplicity we set $m=1$. One needs to make the following choice of
integration constants:
\begin{equation}
   a\equiv a_2=3a_1, \qquad c=0~.
\end{equation}
In this case the expression for the quartic in the numerator of
$\cos^2\zeta$ factorises, with one of the factors cancelling a
quadratic in the denominator. The result is
\begin{equation}
\begin{aligned}
   \me^{6\lambda} &= y^2+a \\
   \cos^2\zeta &= \frac{a-3y^2}{a+y^2}~.
\end{aligned}
\end{equation}
In fact, it is simple to see that the constant $a$ is redundant;
writing $y=\sqrt{a/3}\cos\alpha$ leads to the six-dimensional
metric
\begin{equation}
   \dd s^2_6 = \frac{1}{3} \dd s^2(H^2)
      + \frac{\sin^2\alpha}{3(3+\cos^2\alpha)}\dd s^2 (S^2)
      + \frac{1}{3}\dd\alpha^2
      + \frac{\sin^2\alpha}{3(3+\cos^2\alpha)}
         (\dd\psi+\tilde{P})^2~.
\end{equation}
If we write the metrics on $S^2$ and $H^2$ as
\begin{equation}
\begin{aligned}
   \dd s^2(S^2) &= \dd\theta^2 + \sin^2\theta\dd\nu^2 \\
   \dd s^2(H^2) &= \frac{\dd X^2+\dd Y^2}{Y^2}
\end{aligned}
\end{equation}
then the connection $\tilde{P}$ is given by
\begin{equation}
   \tilde{P} = -\cos\theta \diff \nu -\frac{\diff X}{Y} ~.
\end{equation}

Note that, as usual, the $(\alpha,\psi)$-fibre is a smooth $S^2$
only if we take the period of $\psi$ to be
$2\pi$. However, the full space will then be singular, since the
metric on the base $S^2$ is singular when $\sin\alpha=0$. Remarkably,
the full space can be made smooth by instead choosing the period of
$\psi$ to be $4\pi$. The $(\theta,\nu,\psi)$ part of the
metric is then simply a round three-sphere. Moreover, combined with
the $\alpha$ part we get the metric on a squashed four-sphere.

Thus topologically this solution is different from our previous
examples. The $U(1)$ $\psi$-fibration is now given by
$\mathcal{L}^{1/2}$ instead of $\mathcal{L}$, where $\mathcal{L}$ is
the canonical bundle on $S^2\times H^2$ and the $S^2$ fibre has
conical singularities at $\sin\alpha=0$. As a check one notes that
this choice of period for $\psi$ is still consistent with the
integrated expression~\eqref{chargeom} for $\hat{\Omega}$.

To see the $S^4$ explicitly we introduce the new coordinates
\begin{equation}
   \psi = -(\phi_1 + \phi_2), \qquad
   \nu = \phi_1 - \phi_2
\end{equation}
as well as the constrained variables on a two-sphere
\begin{equation}
   \mu_0=\cos\alpha, \quad
   \mu_1=\sin\alpha\cos\frac{\theta}{2}, \quad
   \mu_2=\sin\alpha\sin\frac{\theta}{2}
\end{equation}
satisfying $\mu_0^2+\mu_1^2+\mu_2^2=1$. The full solution
11-dimensional solution then takes the form
\begin{multline}
\label{MN1}
   \dd s^2_{11} = \Delta^{1/3} \dd s^2_7
      + \frac{1}{4}\Delta^{-2/3}\Bigg\{
         \me^{-4\Phi}\dd\mu_0^2 \\
         + \me^{\Phi}\left[\dd\mu_1^2 + \dd\mu_2^2
            + \mu_1^2\left(\dd\phi_1+\tfrac{1}{2Y}\dd X\right)^2
            + \mu_2^2\left(\dd\phi_2+\tfrac{1}{2Y}\dd X\right)^2
         \right]\Bigg\}
\end{multline}
where
\begin{equation}
   \Delta = \me^{4\Phi}\mu_0^2 + \me^{-\Phi}(\mu_1^2+\mu_2^2)~.
\end{equation}
The constant $\Phi$ is given by $\me^{5\Phi}=4/3$, and the seven-metric
$\dd s^2_7$ is a warped product of $AdS_5$ and $H^2$, given by
\begin{equation}
   \dd s^2_7 = \me^{-8\Phi}\dd s^2{(AdS_5)}_{m=1}
      + \me^{-3\Phi}\frac{\dd X^2+\dd Y^2}{4Y^2}
\end{equation}
where $\dd s^2{(AdS_5)}_{m=1}$ is a unit-radius $AdS_5$-space. The
metric~(\ref{MN1}) is precisely that of the solution, in the form they
presented, found by Maldacena and N\'u\~nez.

\section{Discussion}

Let us summarise what we have found. First, we gave a geometric
formulation of the general conditions that a six-dimensional
manifold $M_6$ must satisfy in order to get a supersymmetric
solution of M-theory of the form of a warped product $AdS_5\times
M_6$. These conditions are summarised in
equations~\eqref{killingvec}--\eqref{dyg}. The four-form flux
is completely determined from the geometry as in (\ref{genflux}).
We found that $M_6$ is
locally the product of a complex four-dimensional space $M_4$ and
a two-dimensional space spanned by a Killing vector $\de/\de\psi$
and an orthogonal direction with coordinate $y$. The complex
structure on $M_4$ is independent of $y$ and $\psi$. The
six-dimensional metric is fixed by a one-parameter family of
K\"ahler metrics on $M_4$ depending on $y$ and a single angular
function $\zeta$, which also fixes the warp factor. The Bianchi
identity and the equations of motion for the four-form are both
implied by the supersymmetry conditions.

We then specialized to the case where $M_6$ is a complex manifold
and the metric is Hermitian, and found a number of new classes of
$AdS_5$ solutions. The natural global structure to consider is
where the $(y,\psi)$ directions form a holomorphic $S^2$ bundle
over the K\"ahler base $M_4$. For compact $M_6$ we found a
complete classification of such geometries, assuming that the
Goldberg conjecture is true. They fall into two classes, where the
base $M_4$ is either (i)~K\"ahler--Einstein (KE) or (ii)~the
product of two constant curvature Riemann surfaces. We found three
families of regular compact solutions: those with positive
curvature KE base (which are classified in \cite{tian,tianyau})
and those with $S^2\times T^2$ or $S^2\times S^2$ base. We also
constructed several families of regular non-compact geometries, as
well as singular geometries.

In the $S^2\times T^2$ case discussed in \ref{tdualsols}
(or the KE case with base $T^4$) one
can reduce on one torus direction to obtain a supersymmetric type
IIA solution which is the direct product of $AdS_5$ with a
five-manifold. If one further T-dualises on the second torus
direction, one gets a supersymmetric type IIB Sasaki-Einstein
solution. This family includes $T^{1,1}/\bZ_2$ as a special case.
A detailed discussion of the new Sasaki-Einstein metrics is
discussed in ref.~\cite{paper2}.
It is perhaps worth stressing that here the T-duality
we implemented preserves supersymmetry, unlike the dualities on the
canonical Sasaki-Einstein $U(1)$ Killing direction of
$S^5$ considered in ref.~\cite{Duff:1998us}, or
$T^{1,1}$~\cite{Dasgupta:1998su}.

These new classes of $AdS_5$ solutions give an array of new
candidates for the supergravity duals of $N=1$ superconformal
field theories. Of primary interest in this context are the
regular compact solutions that we have constructed, but it is
possible that the singular solutions can be resolved in physically
interesting ways. Notice that the general existence of the
$\de/\de\psi$ Killing vector simply reflects the  R-symmetry
of the field theory.

A supergravity solution will serve as a good M/string-theory
background only if the fluxes satisfy appropriate quantisation
conditions. This requirement puts additional constraints on the
classical solutions we have found. This issue will
be analysed in more detail for the solutions of section~\ref{tdualsols}
that are $S^2$ bundles
over $S^2\times T^2$ in~\cite{paper2}, and for this case
flux quantisation reduces the continuous family of
solutions to a countably infinite subclass. In particular,
quantisation of the NS 3-form flux in the type IIA solution is T-dual to the
classical requirement in type IIB that the curvature of a
connection on a $U(1)$ bundle has integral periods (see for
example~\cite{Bouwknegt:2003wp}). On the other
hand, lifting to M-theory, the quantisation of the NS flux implies
that the $G$-flux is also quantised.
In general then, we expect
that these quantisation conditions put constraints on (some of)
the parameters in our solutions. The continuous families of
supergravity solutions that we have found would then only
correspond to true string theory backgrounds at discrete values,
and from the analysis in \cite{paper2}, these may be countably
infinite in number.

It is natural to try to interpret our solutions in $D=11$ in terms
of wrapped or intersecting M5-branes. A very concrete way to make
this connection would be to construct more general solutions with
the property that the solutions presented here arise as
near-horizon limits. For the families of solutions with $S^2\times
S^2 \times T^2$ topology discussed in section~\ref{tdualsols}, and
which have type II dual configurations, a standard interpretation
would be as follows. Consider starting on the type IIB side, where
the solutions arise as the near-horizon limit of $N$ D3-branes
placed at the tip of the Calabi--Yau cone over the
Sasaki--Einstein manifold. Then T-duality along an appropriate
transverse direction maps to a IIA solution which is expected to
arise from a configuration of $N$ D4-branes suspended between
NS5-branes~\cite{Dasgupta:1998su,Uranga:1998vf}. Uplifting to
$D=11$, the M-theory solution is interpreted as the near-horizon
limit of intersecting M5-branes. We expect that, depending on the
choice of parameters, only one of the M-theory, type IIA or type
IIB solutions would provide a good \emph{supergravity} description
of the underlying superconformal field theory. It would be
interesting to see if in any of these limits one can extract the
exact configuration of branes or the nature of the singularity and
hence identify the corresponding conformal field theories.

\subsection*{Acknowledgements}
We thank Alex Buchel, Michael Douglas, Fay Dowker, Jaume Gomis,
Chris Herzog, Ken Intriligator, Rob Myers, Joe Polchinski and
Nemani Suryanarayana for helpful discussions. DM is funded by an
EC Marie Curie Individual Fellowship under contract number
HPMF-CT-2002-01539. JFS is funded by an EPSRC mathematics
fellowship. DW is supported by the Royal Society through a
University Research Fellowship.

\appendix

\section{Conventions}
\label{app:conv}

We use the signature $(-,+,...,+)$. In an orthonormal frame the gamma
matrices satisfy \be
\{\Gamma_\alpha,\Gamma_\beta\}=2\eta_{\alpha\beta} \ee and can be
taken to be real in the Majorana representation. They satisfy, in our
conventions, $\Gamma_{012345678910}=\e_{012345678910}=1$.  Given a
Majorana spinor $\e$ its conjugate  is given by $\bar{\e} =\e^T C$,
where $C$ is the charge conjugation matrix in $D=11$ and satisfies
$C^T=-C$. In the Majorana representation we can choose $C=\Gamma_0$.

The bosonic fields of $D=11$ supergravity consist of a metric, $g$, and
a three-form potential $C$ with four-form field strength $G=\dd C$. The
action for the bosonic fields is given by
\bea
S=\frac{1}{2\kappa^2}\int \diff^{11} x {\sqrt{-g}}R
-\frac{1}{2}G\wedge *G - \frac{1}{6}C\wedge G\wedge G ~.
\eea
The Killing spinor equation is:
\begin{equation}\label{killing}
\nabla_\mu\e+\frac{1}{288}[\Gamma{_\mu}{^{\nu_1\nu_2\nu_3\nu_4}}
-8\delta{_\mu^{\nu_1}}\Gamma^{\nu_2\nu_3\nu_4}]G_{\nu_1\nu_2\nu_3\nu_4}\e=0
\end{equation}
where $\e$ is a Majorana spinor.  The equations of motion are
given by \bea R_{\mu\nu}-\frac{1}{12}(G_{\mu
\s_1\s_2\s_3}G{_{\nu}}{^{\s_1\s_2\s_3}}-
\frac{1}{12}g_{\mu\nu}G^2)&=&0\nn \diff *G +\frac{1}{2}G\wedge  G
&=&0~. \eea
Note that in M-theory, the field-equation for the four-form receives
higher-order gravitational corrections -- in our conventions they can
be found in the appendix of ref.~\cite{Gauntlett:2002fz}.

The Hodge star of a $p$-form $\omega$ is defined by
\be
*\omega_{\mu_1\dots \mu_{11-p}}=\frac{\sqrt
{-g}}{p!}\epsilon_{\mu_1\dots \mu_{11-p}}
{}^{\nu_1\dots\nu_p}\omega_{{\nu_1\dots\nu_p}}~.
\ee

\section{Differential conditions for spinor bilinears}
\label{conds}

We want to manipulate the Killing spinor
equations~\eqref{susycondsone} to define differential conditions on
fermion bilinears on $M_6$. Following a similar calculation
in ref.~\cite{MS}, it is useful to write
\begin{equation}
   \e^+=\tfrac{1}{\sqrt 2}\,\xi \qquad
   \e^-=-\tfrac{1}{\sqrt 2}\,\ii\gamma_7\xi~.
\end{equation}
We then get
\bea\label{gravitino-like}
   \nabla_m\e^\pm \mp
      \frac{m}{2}\gamma_m\e^\mp \mp \frac{1}{24}\gamma^{n_1n_2n_3}
      \me^{-3\lambda}G_{mn_1n_2n_3}\e^\pm &=& 0\nn
   \gamma^m\nabla_m\lambda\e^\pm \pm m \e^\mp \pm
      \frac{1}{144}\gamma^{m_1m_2m_3m_4}
      \me^{-3\lambda}G_{m_1m_2m_3m_4}\e^\pm & = & 0~.
\eea
Given that $\gamma_m^\dagger = \gamma_m$, one can derive some useful
identities
\bea
   \frac{1}{144}\me^{-3\lambda} G_{pqrs} \bar\e^{\pm }
      [\gamma^{pqrs},A]_- \e^\pm \pm \nabla_m \lambda \,\bar\e^{\pm}
      [\gamma^{m},A]_- \e^\pm + m (\bar\e^{\mp }A\e^\pm - \bar\e^{\pm }A
      \e^\mp) = 0\nn
   \frac{1}{144} \me^{-3\lambda}G_{pqrs} \bar\e^{\pm} [\gamma^{pqrs},A]_+
      \e^\pm \pm \nabla_m \lambda \,\bar\e^{\pm} [\gamma^{m},A]_+\e^\pm + m
      (\bar\e^{\mp }A\e^\pm + \bar\e^{\pm }A \e^\mp) = 0\nn
   \frac{1}{144}\me^{-3\lambda} G_{pqrs} \bar\e^{+ } [\gamma^{pqrs},A]_\pm
   \e^- +\nabla_m \lambda \,\bar\e^{+ } [\gamma^{m},A]_\mp\e^- + m
   (\bar\e^{-}A\e^- \pm \bar\e^{+}A \e^+)  = 0\label{lastuseful}
\eea
where $[\cdot,\cdot]_\pm$ refer to anticommutator or commutator,
$A$ is a general Clifford matrix, and  $\bar\e\equiv \e^\dagger$.
There are similar formulae involving $\e^{\pm\trsp}$.

Let us now consider the bilinears that can be constructed from
$\e^\pm$.  We first analyse the scalars. By definition
$\bar\e^+\e^+=\bar\e^-\e^-=\frac{1}{2}\bar{\xi}\xi$. Using
(\ref{gravitino-like}) we find $\nabla(\bar\e^{+}\e^+)=0$.  Thus
we can normalise the spinor so that
\bea\label{normspin}
   \bar\e^+\e^+ = \bar\e^-\e^- = 1~.
\eea
On the other hand $\nabla(\bar\e^{+}\e^-) \neq 0$, and we
parameterise this non-trivial function, which takes values in the
interval $[-1,1]$, as
\bea\label{anglespin}
   \bar\e^{+}\e^- \equiv \sin \zeta ~.
\eea
Of the other possible scalars, notice that $\e^{+\trsp} \e^- = 0$
while $\e^{+\trsp} \e^+ = - \e^{-\trsp} \e^-\equiv f$ is {\it a priori}
non-zero.

We can also construct the following tensor fields as bilinears:
\bea\label{bilindef}
   \tilde{K}^1_m  &=&   \bar\e^+ \gamma_m \e^+\nn
   \tilde{K}^2_m &=& i \bar\e^+ \gamma_m \e^-\nn Y_{mn} & = & -i
   \bar\e^+ \gamma_{mn} \e^+\nn Y'_{mn} & = & i \bar\e^+ \gamma_{mn}\e^-\nn
   \tilde\Omega_{mn} & = & \e^{+\trsp} \gamma_{mn} \e^-\nn
   X_{mnp} & = & \e^{+\trsp} \gamma_{mnp} \e^+\nn V_{mnp} & = & \bar\e^+
   \gamma_{mnp} \e^-~.
\eea
Consideration of other bilinears turns out to be redundant and we
will not include them in the following analysis.

Observe that the covariant derivative of $\tilde K^2$ is given by: 
\bea\label{k2kill}
\nabla_{m} \tilde K^2_n  & = &
-\frac{1}{4}\me^{-3\lambda}G_{mnpq} Y'^{\,pq} -m  Y_{mn} ~.
\eea
This implies that $\nabla_{(m} \tilde K^2_{r)}=0$ and therefore
$(\tilde K^2)^m$ are the components of a Killing vector.  Next,
setting $A=1$ in the last equation of \p{lastuseful} with the lower
sign immediately gives the useful condition
\bea
{\cal L}_{\tilde K^2} \lambda & = & 0 \label{warpok}~.
\eea
Conditions on the exterior derivatives of all the bilinears can be
obtained, and we find:
\bea
   \diff (\me^{3\lambda} f) & = & 0\label{ns1}\\
   \me^{-3\lambda}\diff(\me^{3\lambda} \sin\zeta) & = & 2 m \tilde{K}^1\label{ns2}\\
   \diff (\me^{3\lambda} \tilde{K}^1) & = & 0\label{ns3}\\
   \me^{-6\lambda}\diff (\me^{6\lambda} \tilde{K}^2 ) & = &
      \me^{-3\lambda} * G + 4 m Y\label{ns6}\\
   \diff (\me^{6\lambda} Y ) & = & 0\label{ns4}\\
   \me^{-6\lambda}\diff (\me^{6\lambda} \tilde\Omega ) &= & 3m \, X\label{ns5}\\
   \me^{-6\lambda}\diff (\me^{6\lambda}X) & = & - f\,\me^{-3\lambda} G\label{ns7}\\
   \me^{-6\lambda}\diff (\me^{6\lambda}V) & = &
      \me^{-3\lambda}G \sin\zeta +2 m * Y'~.\label{ns8}
\eea
Notice that, when $m$ is non-zero, which is the focus of this
paper, (\ref{ns3}) is implied by (\ref{ns2}), and (\ref{ns4}) is
implied by (\ref{ns6}) and the $G$ equation of motion. Moreover,
(\ref{ns5}) together with (\ref{ns7}) imply the important fact
that
\begin{equation}
\label{fcond}
   f \equiv \e^{+\trsp}\e^+ = \tfrac{1}{2}{\xi}^{\trsp}\xi = 0~.
\end{equation}

\section{$SU(3)$ and $SU(2)$ structures in $d=6$}
\label{su2structure}

We now rewrite the conditions on the spinor bilinears derived in
the appendix~\ref{conds} in terms of an explicit local $SU(2)$
structure. We start by defining $SU(3)$ and $SU(2)$ structures in
$d=6$.  We work in a basis in which the gamma-matrices are
imaginary. A single unit norm chiral spinor in $d=6$, satisfying
\bea \bar\eta_1\eta&=&1\nn -\mathrm{i} \gamma_7 \eta &=& \eta \eea
defines an $SU(3)$ structure with two-form $j$ and $(3,0)$-form
$\omega$ given by
\bea\label{d=6su3}
   j=-i\bar\eta_1\gamma_{(2)}\eta_1\nn
   \omega=\eta_1^T\gamma_{(3)}\eta_1
\eea
where $\gamma_{(n)}$ is the $n$-form $\frac{1}{n!}\gamma_{i_1\dots
i_n}e^{i_1}\wedge\dots\wedge e^{i_n}$. For further discussion, see
for example ref.~\cite{intrinsic}.

Now consider two orthogonal unit norm chiral spinors $\eta_1$,
$\eta_2$  satisfying
\bea
   \bar \eta_1 \eta_1 &=&1\nn
   \bar \eta_2 \eta_2 &=&1\nn
   \bar \eta_1 \eta_2&=&0
\eea
and $ -\mathrm{i} \gamma_7 \eta_a = \eta_a$.  Together these define
two canonical $SU(3)$ structures, specified by forms $(j_1,\omega_1)$
and $(j_2,\omega_2)$ defined as in \p{d=6su3}. Equivalently,
$\eta_1$, $\eta_2$  define a canonical $SU(2)$ structure in $d=6$.
Such a structure is specified by two one-forms $K^1, K^2$ and three
two-forms $J^m$ defined by
\bea\label{defplus}
   J^m&=&-\frac{i}{2}\sigma_m^{\a\b}\bar\eta_\a\gamma_{(2)}\eta_\b\nn
   K^1-iK^2&=&-\frac{1}{2}\epsilon^{\a\b}\eta_\a^T\gamma_{(1)}\eta_\b
\eea
where $\sigma_m$ are Pauli matrices. The $d=6$ metric has the form
\be
   \dd s^2=e^ie^i+(K^1)^2+(K^2)^2
\ee
with $J^m=(1/2)J^m_{ij}e^{i}\wedge e^j$ satisfying the algebraic
identities of a $d=4$ $SU(2)$ structure. We also define
\bea\label{defplustwo}
   \Omega &\equiv& J^2+iJ^1\nn J&\equiv& J^3
\eea
so the $SU(2)$ structure in $d=6$ is equivalently specified by
$(J,\Omega, K^1, K^2)$. We naturally define
$\vol_6=\vol_4\wedge K^1\wedge K^2$ where
$\vol_4=\frac{1}{2}J\wedge J$. In addition we have that
each $J^i$ is self-dual with respect to $\vol_4$ and also,
after raising an index, we have $J^1\cdot J^2 =J^3$.

For completeness, we note that the two $SU(3)$ structures defined
by~\p{d=6su3} can be written in terms of the $d=6$ $SU(2)$ structure
as
\bea
   j_1&=& J-K^1\wedge K^2, \qquad
   \omega_1=-\Omega\wedge(K^1-iK^2)\nn
   j_2&=& -J-K^1\wedge K^2, \qquad
   \omega_2=-\bar\Omega\wedge (K^1-iK^2)~.
\eea

In doing calculations it is often easiest to specify the spinors
in terms of some concrete convenient projections. In particular,
this provides a simple derivation of the above formulae. For
example, for the spinor $\eta_1$ we take
\bea
   \gamma_{12}\eta_1=\gamma_{34}\eta_1=-\gamma_{56}\eta_1=i\eta_1\nn
   \gamma_{135}\eta_1=-\eta_1^*~.
\eea
For the second spinor $\eta_2$ we take
\bea
   -\gamma_{12}\eta_2=-\gamma_{34}\eta_2=-\gamma_{56}\eta_2=i\eta_2\nn
   \gamma_{135}\eta_2=-\eta_2^*~.
\eea
In addition, the two spinors are related by
\be
   \gamma_5\eta_2^*=\eta_1~.
\ee
This then leads to $J^1=e^{14}+e^{23}$, $J^2=e^{13}-e^{24}$,
$J^3=e^{12}+e^{34}$, $K^1=e^5$ and $K^2=e^6$. Note that the $J^m$
are indeed self-dual and that $J^1\cdot J^2 =J^3$.

We would now like to relate the non-chiral spinor $\xi$ entering
the Killing spinor equations~\eqref{susycondsone} to a canonical
$d=6$ $SU(2)$ structure, as defined above. We first observe that
we can always take two non-orthogonal unit-norm chiral spinors to
be $\eta_1$ and $a\eta_1 +b \eta_2$ with $|a|^2+|b|^2=1$.
Therefore, a general non-chiral spinor $\xi$ can be decomposed as
\be
   \xi\equiv\xi_+ + \xi_-=f_1\eta_1 +f_2(a\eta_1 +b \eta_2)^*
\ee
where $\eta_i$ are uniquely defined up to phases. We fix this 
phase freedom\footnote{To see this, note that two chiral spinors
  $\xi_i$, $i=1,2$ define four real scalars: $\bar\xi_i\xi_i\equiv f_i^2$ and 
$\bar \xi_1\xi_2\equiv f_1 f_2 a$. One can then define $\eta_1=\xi_1/f_1$ 
and $\eta_2=(1-|a|^2)^{-1/2} (\xi_2/f_2-a\xi_1/f_1)$.}
by taking 
$f_i$ real and $b=(1-|a|^2)^{1/2}$. Now,
since our spinor satisfies \p{normspin}, we set $f_1 ={\sqrt 2}
\cos\alpha$ and $f_2={\sqrt 2} \sin\alpha$. Moreover, since we
found that $f$ was zero~\eqref{fcond}, we conclude that we can
choose $a=0$ and hence $b=1$. Hence we have
\bea
   \xi_+&=&{\sqrt 2} \cos\alpha\eta_1\nn
   \xi_-&=&{\sqrt 2} \sin\alpha\eta_2^*
\eea
and consistency with \p{anglespin} implies $\cos 2\alpha=\sin\zeta$.
Note that this implies $\cos\zeta=\sin 2\alpha$ where we have fixed
an arbitrary sign.

Having established how our spinor $\xi$ is related to a
canonically defined $SU(2)$ structure defined by $\eta_1$,
$\eta_2$, we can now translate the bilinears defined
in~\p{bilindef} into those naturally defined by the $SU(2)$
structure. We find:
\bea
   \tilde{K}^1 & = & K^1\,\cos\zeta\nn
   \tilde{K}^2 & = & K^2\,\cos\zeta\nn
   Y & = & J - K^1 \wedge K^2\,\sin\zeta \nn
   Y' & = & -J \sin\zeta  + K^1 \wedge K^2 \nn
   \tilde\Omega & = & \Omega \cos\zeta\nn
   X & = & \Omega \wedge (-K_1 \sin\zeta + i  K^2)\nn
   V & = & J \wedge K^2 \cos\zeta
\eea
together with
\be
   \vol_6 = \frac{1}{2}J\wedge J \wedge K^1 \wedge K^2~.
\ee
We then find that~\eqref{ns1}--\eqref{ns8} are equivalent to
\begin{equation}
\begin{aligned}
   \me^{-3\lambda}\diff(\me^{3\lambda}\sin\zeta) & =
      2 m K^1\cos\zeta \\
   \me^{-6\lambda}\diff (\me^{6\lambda} \Omega\cos\zeta ) & =
      3 m \, \Omega \wedge (-K^1 \sin\zeta + i  K^2) \\
   \me^{-6\lambda}\diff(\me^{6\lambda} K^2 \cos\zeta) & =
      \me^{-3\lambda}* G + 4 m (J - K^1\wedge K^2\,\sin\zeta) \\
   \me^{-6\lambda}\diff (\me^{6\lambda}J \wedge K^2 \cos\zeta) & =
      \me^{-3\lambda}G \sin\zeta
      + m(J \wedge J - 2 J \wedge K^1 \wedge K^2 \sin\zeta)
\end{aligned}
\end{equation}
where in the last formula we have used $*J=J\wedge K^1 \wedge K^2$.

\section{Minkowski$_5$ solutions}
\label{m=0}

Here we analyse the supersymmetry constraints on the geometry in
the case of vanishing five-dimensional cosmological constant,
$m=0$. This corresponds to the most general supersymmetric
solutions of $D=11$ supergravity that are warped products of
Minkowski$_5$-space with a six-dimensional manifold. It has been
argued in ref.~\cite{gukov}, using an argument based on an
effective superpotential, that it is not possible to have such
supersymmetric M-theory vacua with non-trivial fluxes. Our results
show that these are possible, if we consider a suitably general
spinor ansatz. The reason for this apparent discrepancy is that
the argument of ref.~\cite{gukov} implicitly assumed that the
internal manifold is Calabi--Yau, which indeed turns out to be
incompatible with flux.

Let us now turn to the detailed analysis of the supersymmetry
constraints. The basic conditions are given by the equations (\ref
{ns1})--(\ref {ns8}),  upon setting $m=0$. From (\ref{ns3}) we
deduce that we have an integrable almost-product structure. Next,
from (\ref{k2kill}) (which must in fact follow from (\ref
{ns1})--(\ref {ns8})), we see that $\tilde {K}^2=\cos\zeta K^2$ is a 
Killing vector. Moreover, we also have
$\mathcal{L}_{\tilde{K}^2}\lambda=\mathcal{L}_{\tilde{K}^2}\zeta=
\mathcal{L}_{\tilde{K}^2}K^1= 0$. Therefore, similar to the $m\ne
0$ case, the $d=6$ metric can locally be written as
\bea
\dd s^2 & = & g^4_{ij}(x,y) \dd x^i \dd x^j +
\me^{-6\lambda}\sec^2\zeta \dd y^2 +\cos^2\zeta (\dd\psi + \rho)^2
\eea
with
\bea
K^1 = \me^{-3\lambda} \sec\zeta\diff y \qquad \qquad K^2=
\cos\zeta ( \diff \psi+\rho)
\eea
and $\partial_\psi\lambda=\partial_\psi\zeta=0$.
Using  equation  (\ref {ns6}), we get the following expression for the
flux:
\bea
G & = & \de_y (\me^{6\lambda}\cos^2\zeta) \mathrm{vol}_4-
\me^{-3\lambda}\sec\zeta *_4 \diff_4 (\me^{6\lambda}\cos^2\zeta)
\wedge K^1 - \cos^3 \zeta \me^{6\lambda} *_4 \de_y \rho \wedge K^2\nn
&&  + \me^{3\lambda}\cos^2 \zeta *_4 \diff_4 \rho \wedge K^1 \wedge K^2~.
\label{genfluxm0}
\eea
Notice this expression is slightly different from that one would
naively obtain by setting $m=0$ in (\ref {genflux}).

Now from (\ref{ns2}) we see that there are two cases to consider
separately, depending on whether $\sin\zeta$ vanishes or not (note
that $\sin\zeta=0$ is not possible when $m\ne 0$).

\subsection{$\sin\zeta=0$}

For this case we have $Y'=K^1\wedge K^2$, so
taking the antisymmetric part of (\ref{k2kill}) we get
\bea
\diff \rho & = & -\me^{-3\lambda} (K^1\wedge K^2) \lrcorner \, G~.
\label{nosinz}
\eea
Substituting  (\ref{genfluxm0}) into (\ref {nosinz}) we
immediately conclude that $\de_y \rho=0$ and
\bea (\diff_4 \rho)^+ & = & 0~.  \eea Notice that in the present
case the chiral spinors $\xi_\pm$ have constant norm, and hence
they define a {\em global} six-dimensional $SU(2)$-structure.
Moreover, we find \bea \diff (\me^{6\lambda}J) & = & 0\nn \diff
(\me^{6\lambda}\Omega) &  = &  0 \eea implying that the
four-dimensional slices, at fixed $y$, are conformal to
$y$-independent hyper-K\"ahler manifolds. Therefore the general
form of the metric is \bea \dd s^2 = (\dd \psi + \rho)^2+
\me^{-6\lambda} [\dd\hat{s}_4^2(\mathrm{HK}) + \dd y^2]~. \eea It
is a simple matter to check that (\ref{ns8}) is trivially
satisfied, while (\ref {ns7}) forces $f=0$. The $G$-flux reads
\bea
G & = & - \de_y (\me^{-6\lambda}) \hat {\mathrm{vol}}_4 +  \hat{*}_4 \diff_4 (\me^{-6\lambda})\wedge \diff y - \diff_4  \rho \wedge \diff y \wedge (\diff \psi +\rho)
\eea
and the (source-less) Bianchi identity $\diff G=0$ implies
\bea
\hat\Box\, \me^{-6\lambda} + \de_y^2 \me^{-6\lambda} & =
& -\frac{1}{2} ||\hat{\diff_4 \rho} ||^2
\eea
which must be solved in order to complete the solution.

It is interesting to note that the geometries we have found arise
when a fivebrane is transverse to a hyper-K\"ahler manifold times
a flat direction, with a single flat longitudinal direction of the
fivebrane fibred over the hyper-K\"ahler manifold. When $\de/\de y
$ is a Killing vector, we can reduce to type IIA theory and
recover the results of ref.~\cite{intrinsic}, where the
corresponding geometry is related to wrapped NS5-branes.  Examples
of these solutions appeared in ref.~\cite{Lu}, where the
four-space was taken to be Eguchi--Hanson\footnote{Note that the
construction of ref.~\cite{Lu} has been generalized in
ref.~\cite{intrinsic} to an arbitrary number of twisted
directions.  For instance, twisted $T^2$ bundles over a
hyper--K\"ahler four-manifold yield potentially interesting flux
compactifications to four dimensions.}.

Alternatively, we can also make a Kaluza--Klein reduction along the
$\psi$-direction.  We thus obtain supersymmetric IIA solutions,
with constant dilaton and non-trivial RR two-form and four-form,
plus NS $B$-field $B^{NS}= - y \, \diff_4 \rho$. Finally, we note
that simply setting $\rho=0$ we recover the conditions on the
local geometry obtained by Witten~\cite{witten} for M-theory
compactifications to six-dimensional Minkowski space.

\subsection{$\sin\zeta\neq 0$}

We must now have $\me^{3\lambda}\sin\zeta=c$ for some constant
$c$. Hence, from (\ref{ns8}), we see that $G$ is exact:
\be
c\,G=\dd(\me^{6\lambda}\cos\zeta J\wedge K^2)
\ee
so that the Bianchi identity is automatically satisfied (as expected
from the argument in section 2.2).  The conditions on the
base geometry can be conveniently expressed in terms of rescaled
quantities, $\hat g_{ij}=\me^{6\lambda} g_{ij}$, $\hat J=\me^{6\lambda} J$ and $\hat \Omega=\me^{6\lambda} \Omega$.
We find
\bea
\diff_4 \hat{J} & = & 0 \nn \diff_4 (\cos\zeta \hat{\Omega}) & =
& 0
\label{acym0}
\eea
which state that the base manifold, at fixed $y$, is
almost-Calabi--Yau. In addition, we have
\bea
\de_y \hat J&=& -c\,\diff_4 \rho \nn \de_y
(\cos\zeta\hat\Omega) & = & 0~. \label{basem0}
\eea
Comparing the expressions for the flux computed from
(\ref{genfluxm0}) and (\ref{ns8}), we obtain an additional
constraint:
\bea c*_4 \de_y \rho & = & J \wedge \diff_4 \sec^2\zeta~.
\label{extram0} \eea
Note that here we can derive an expression for $J$ that is similar
to that in the $m\neq 0$ case, which reads
\bea (\diff_4 \rho)^+ & = & \frac{-c}{\sin\zeta\cos\zeta}\de_y \zeta
J~.  \eea
It is now a simple matter to deal with the remaining conditions
(\ref{ns7}) and (\ref{ns1}). Using the equations
(\ref{acym0})--(\ref{basem0}), we find that (\ref{ns7}) reduces to 
\bea f \me^{-3\lambda} G & = & -\cos\zeta \diff y \wedge \Omega
\wedge [i\de_y \rho +\frac{1}{c}\diff_4 \sec^2\zeta]~. \eea
However, upon using (\ref{extram0}), one shows that the right-hand
side vanishes.  Hence $f=0$ in the present case also. Observe that
an integrability condition between the first equation in
(\ref{basem0}) and (\ref{extram0}) yields the following
second-order equation
\bea
\de_y^2 (\me^{6\lambda}{J}) -2ci \de\bar\de \sec^2\zeta & = & 0~.
\eea

We conclude by presenting a simple example of a (singular)
solution in this class. Consider the ansatz for the
four-dimensional base metric
\bea
\dd\hat{s}^2 & = & \Lambda (y) \dd\hat{s}^2_0
\eea
where the re-scaled metric is $y$-independent and hyper--K\"ahler,
with associated hyper-K\"ahler structure
$\hat{J}_0,\hat{\Omega}_0$. Setting $c=1$ for simplicity, it is
straightforward to check that all the conditions are satisfied,
provided
\bea \Lambda & = & \sec\zeta = 1 + y\nn \diff_4 \rho & = &
\hat{J}_0\nn \de_y \rho & = & 0~.  \eea
The corresponding flux reads
\bea G & = & \diff \left[ \frac{1}{(1+y)} \hat{J}_0\wedge (\diff
\psi + \rho) \right] \eea
and the six-dimensional metric is
\bea \dd s^2_6 & = &
\frac{y(2+y)}{1+y}\left[ \dd\hat{s}^2_0 + (1+y)\dd y^2\right] +
\frac{1}{(1+y)^2} (\dd\psi + \rho)~.
\eea
It is interesting to note
that in the limit $y\to \infty$ ($\cos\zeta \to 0$) the flux
vanishes and, correspondingly, the eleven-dimensional metric
asymptotes to
\bea \dd s^2 & = & \dd s^2 (\mathrm{Mink}_5) + y \dd\hat{s}^2_0 + y^2 \dd y^2
+ \frac{1}{y^2} (\dd\psi + \rho)^2~. \eea
It is straightforward to check that the internal metric is indeed
Calabi--Yau. This behaviour, in fact, is not accidental. In
general, when $\cos\zeta \to 0$, the warp factor will become
asymptotically constant\footnote{Sub-leading behaviour could be
relevant.}, and in this case, from the Einstein equation, it
follows that the flux must vanish (when $m=0$). Therefore, this
interpolating feature is likely to be quite generic.
Unfortunately, as $y\to 0$ the metric develops a singularity.



\begin{thebibliography}{99}


\bibitem{mal} J.~M.~Maldacena, ``The large $N$ limit of superconformal
field theories and supergravity,'' Adv.\ Theor.\ Math.\ Phys.\  {\bf
2} (1998) 231 [Int.\ J.\ Theor.\ Phys.\  {\bf 38} (1999) 1113]
[arXiv:hep-th/9711200].

\bibitem{Klebanov:1998hh} I.~R.~Klebanov and E.~Witten,
``Superconformal field theory on threebranes at a Calabi--Yau
singularity,'' Nucl.\ Phys.\ B {\bf 536} (1998) 199
[arXiv:hep-th/9807080].


\bibitem{Acharya:1998db} B.~S.~Acharya, J.~M.~Figueroa-O'Farrill,
C.~M.~Hull and B.~Spence, ``Branes at conical singularities and
holography,'' Adv.\ Theor.\ Math.\ Phys.\  {\bf 2} (1999) 1249
[arXiv:hep-th/9808014].


\bibitem{Morrison:1998cs} D.~R.~Morrison and M.~R.~Plesser,
``Non-spherical horizons. I,'' Adv.\ Theor.\ Math.\ Phys.\  {\bf 3}
(1999) 1 [arXiv:hep-th/9810201].


\bibitem{Pope:1988xj} C.~N.~Pope and P.~van Nieuwenhuizen,
``Compactifications Of $D=11$ Supergravity On K\"{a}hler Manifolds,''
Commun.\ Math.\ Phys.\  {\bf 122} (1989) 281.


\bibitem{Gauntlett:2002rv} J.~P.~Gauntlett, N.~Kim, S.~Pakis and
D.~Waldram, ``M-theory solutions with AdS factors,'' Class.\ Quant.\
Grav.\  {\bf 19} (2002) 3927 [arXiv:hep-th/0202184].


\bibitem{Cvetic:2000cj} M.~Cvetic, H.~Lu, C.~N.~Pope and
J.~F.~Vazquez-Poritz, ``AdS in warped spacetimes,'' Phys.\ Rev.\ D
{\bf 62} (2000) 122003 [arXiv:hep-th/0005246].

\bibitem{Alishahiha:1999ds} M.~Alishahiha and Y.~Oz, ``AdS/CFT and BPS
strings in four dimensions,'' Phys.\ Lett.\ B {\bf 465} (1999) 136
[arXiv:hep-th/9907206].


\bibitem{Fayyazuddin:1999zu} A.~Fayyazuddin and D.~J.~Smith,
``Localized intersections of M5-branes and four-dimensional
superconformal field theories,'' JHEP {\bf 9904} (1999) 030
[arXiv:hep-th/9902210].


\bibitem{malnun} J.~M.~Maldacena and C.~N\'u\~nez, ``Supergravity
description of field theories on curved manifolds and a no  go
theorem,'' Int.\ J.\ Mod.\ Phys.\ A {\bf 16} (2001) 822
[arXiv:hep-th/0007018].


\bibitem{Cucu:2003bm} S.~Cucu, H.~Lu and J.~F.~Vazquez-Poritz, ``A
supersymmetric and smooth compactification of M-theory to $AdS_5$,''
Phys.\ Lett.\ B {\bf 568} (2003) 261 [arXiv:hep-th/0303211].

\bibitem{Fayyazuddin:2000em} A.~Fayyazuddin and D.~J.~Smith, ``Warped
AdS near-horizon geometry of completely localized intersections  of
M5-branes,'' JHEP {\bf 0010} (2000) 023 [arXiv:hep-th/0006060].

\bibitem{Gauntlett:2002sc} J.~P.~Gauntlett, D.~Martelli, S.~Pakis and
D.~Waldram, ``$G$-structures and wrapped NS5-branes,'' to appear
in Comm. Math. Phys. [arXiv:hep-th/0205050].


\bibitem{Friedrich:2001nh}
T.~Friedrich and S.~Ivanov,
``Parallel spinors and connections with skew-symmetric torsion in string
theory,''
Asian J. Math.  {\bf 6}  (2002), no. 2, 303--335
[arXiv:math.dg/0102142].

\bibitem{Gauntlett:2001ur}
J.~P.~Gauntlett, N.~Kim, D.~Martelli and D.~Waldram,
``Fivebranes wrapped on SLAG three-cycles and related geometry,''
JHEP {\bf 0111} (2001) 018
[arXiv:hep-th/0110034].

\bibitem{Ivanov:2001ma}
S.~Ivanov,
``Connection with torsion, parallel spinors and geometry of Spin(7)
manifolds,''
arXiv:math.dg/0111216.


\bibitem{Friedrich:2001yp}
T.~Friedrich and S.~Ivanov,
``Killing spinor equations in dimension 7 and geometry of integrable
$G_2$-manifolds,''
J. Geom. Phys.  {\bf 48}  (2003), no. 1, 1--11,
[arXiv:math.dg/0112201].

\bibitem{Gurrieri:2002wz}
S.~Gurrieri, J.~Louis, A.~Micu and D.~Waldram,
``Mirror symmetry in generalized Calabi--Yau compactifications,''
Nucl.\ Phys.\ B {\bf 654} (2003) 61
[arXiv:hep-th/0211102].

\bibitem{Cardoso:2002hd}
G.~L.~Cardoso, G.~Curio, G.~Dall'Agata, D.~Lust, P.~Manousselis and G.~Zoupanos,
``Non-K\"{a}hler string backgrounds and their five torsion classes,''
Nucl.\ Phys.\ B {\bf 652} (2003) 5
[arXiv:hep-th/0211118].

\bibitem{Gauntlett:2002fz} J.~P.~Gauntlett and S.~Pakis, ``The
geometry of $D=11$ Killing spinors,'' JHEP {\bf 0304} (2003) 039
[arXiv:hep-th/0212008].

\bibitem{intrinsic} J.~P.~Gauntlett, D.~Martelli and D.~Waldram,
``Superstrings with intrinsic torsion,'' 
Phys.\ Rev.\ D {\bf 69}, 086002 (2004)
[arXiv:hep-th/0302158].

\bibitem{kmt} P.~Kaste, R.~Minasian and A.~Tomasiello,
``Supersymmetric M-theory compactifications with fluxes on
seven-manifolds and $G$-structures,'' JHEP {\bf 0307} (2003) 004
[arXiv:hep-th/0303127].

\bibitem{MS} D.~Martelli and J.~Sparks, ``$G$-structures, fluxes and
calibrations in M-theory,'' Phys.\ Rev.\ D {\bf 68} (2003) 085014
[arXiv:hep-th/0306225].

\bibitem{Behrndt:2003ih}
K.~Behrndt and M.~Cvetic,
``Supersymmetric intersecting D6-branes and fluxes in massive type IIA string
theory,''
Nucl.\ Phys.\ B {\bf 676}, 149 (2004)
[arXiv:hep-th/0308045].

\bibitem{Gauntlett:2003wb}
J.~P.~Gauntlett, J.~B.~Gutowski and S.~Pakis,
``The geometry of $D=11$ null Killing spinors,''
JHEP {\bf 0312}, 049 (2003)
[arXiv:hep-th/0311112].

\bibitem{Behrndt:2003zg}
K.~Behrndt and C.~Jeschek,
``Fluxes in M-theory on 7-manifolds: $G$-structures and superpotential,''
arXiv:hep-th/0311119.

\bibitem{Fidanza:2003zi}
S.~Fidanza, R.~Minasian and A.~Tomasiello,
``Mirror symmetric SU(3)-structure manifolds with NS fluxes,''
arXiv:hep-th/0311122.

\bibitem{Dall'Agata:2003ir} G.~Dall'Agata and N.~Prezas, ``$\mathcal{N}=1$
geometries for M-theory and type IIA strings with fluxes,''
arXiv:hep-th/0311146.

\bibitem{Ivanov:2003nd}
P.~Ivanov and S.~Ivanov,
``SU(3)-instantons and $G_2$, Spin(7)-heterotic string solitons,''
arXiv:math.dg/0312094.


\bibitem{Gauntlett:2002nw}
J.~P.~Gauntlett, J.~B.~Gutowski, C.~M.~Hull, S.~Pakis and H.~S.~Reall,
``All supersymmetric solutions of minimal supergravity in five dimensions,''
Class.\ Quant.\ Grav.\  {\bf 20}, 4587 (2003)
[arXiv:hep-th/0209114].

\bibitem{Gauntlett:2003fk}
J.~P.~Gauntlett and J.~B.~Gutowski,
``All supersymmetric solutions of minimal gauged supergravity in five
dimensions,''
Phys.\ Rev.\ D {\bf 68}, 105009 (2003)
[arXiv:hep-th/0304064].

\bibitem{Gutowski:2003rg}
J.~B.~Gutowski, D.~Martelli and H.~S.~Reall,
``All supersymmetric solutions of minimal supergravity in six dimensions,''
Class.\ Quant.\ Grav.\  {\bf 20}, 5049 (2003)
[arXiv:hep-th/0306235].

\bibitem{Caldarelli:2003pb}
M.~M.~Caldarelli and D.~Klemm,
``All supersymmetric solutions of $\mathcal{N}=2$, $D=4$ gauged supergravity,''
JHEP {\bf 0309}, 019 (2003)
[arXiv:hep-th/0307022].


\bibitem{Cariglia:2004kk}
M.~Cariglia and O.~A.~P.~Conamhna,
``The general form of supersymmetric solutions of $\mathcal{N}=(1,0)$
U(1) and SU(2) gauged supergravities in six dimensions,''
arXiv:hep-th/0402055.



\bibitem{apostolov} V. Apostolov, T. Draghici, and A. Moroianu, ``A
splitting theorem for K\"{a}hler manifolds whose Ricci tensors have
constant eigenvalues'', math.DG/0007122.

\bibitem{tian}
G. Tian,``On K\"ahler-Einstein metrics on certain K\"ahler manifolds with
$c_1(M)>0$'', Invent. Math. {\bf 89} (1987) 225-246.

\bibitem{tianyau}
G. Tian and S.T.~Yau,
``On K\"ahler-Einstein metrics on complex surfaces with $C_1>0$'',
Commun. Math. Phys. {\bf 112} (1987) 175-203.

\bibitem{paper2}
J.~P.~Gauntlett, D.~Martelli, J.~Sparks and D.~Waldram,
``Sasaki-Einstein metrics on $S^2\times S^3$,''
arXiv:hep-th/0403002.

\bibitem{Becker:1996gj} K.~Becker and M.~Becker, ``M-Theory on
Eight-Manifolds,'' Nucl.\ Phys.\ B {\bf 477} (1996) 155
[arXiv:hep-th/9605053].

\bibitem{hkpap} J.~P.~Gauntlett, G.~W.~Gibbons, G.~Papadopoulos and
P.~K.~Townsend, ``Hyper-K\"{a}hler manifolds and multiply intersecting
branes,'' Nucl.\ Phys.\ B {\bf 500} (1997) 133 [arXiv:hep-th/9702202].

\bibitem{Bergshoeff:1995as} E.~Bergshoeff, C.~M.~Hull and T.~Ortin,
``Duality in the type II superstring effective action,'' Nucl.\ Phys.\
B {\bf 451} (1995) 547 [arXiv:hep-th/9504081].

\bibitem{Duff:1998us} M.~J.~Duff, H.~Lu and C.~N.~Pope,
``$AdS_5\times S_5$ untwisted,'' Nucl.\ Phys.\ B {\bf 532} (1998) 181
[arXiv:hep-th/9803061].


\bibitem{Dasgupta:1998su}
K.~Dasgupta and S.~Mukhi,
``Brane constructions, conifolds and M-theory,''
Nucl.\ Phys.\ B {\bf 551}, 204 (1999)
[arXiv:hep-th/9811139].


\bibitem{Bouwknegt:2003wp}
P.~Bouwknegt, J.~Evslin and V.~Mathai,
``On the topology and $H$-flux of T-dual manifolds,''
arXiv:hep-th/0312052.


\bibitem{Uranga:1998vf}
A.~M.~Uranga,
JHEP {\bf 9901} (1999) 022
[arXiv:hep-th/9811004].



\bibitem{gukov} K.~Behrndt and S.~Gukov, ``Domain walls and
superpotentials from M theory on Calabi--Yau  three-folds,'' Nucl.\
Phys.\ B {\bf 580}, 225 (2000) [arXiv:hep-th/0001082].


\bibitem{Lu} H.~Lu and J.~F.~Vazquez-Poritz, ``Resolution of
overlapping branes,'' Phys.\ Lett.\ B {\bf 534}, 155 (2002)
[arXiv:hep-th/0202075].

\bibitem{witten} E.~Witten, ``Strong Coupling Expansion Of Calabi--Yau
Compactification,'' Nucl.\ Phys.\ B {\bf 471} (1996) 135
[arXiv:hep-th/9602070].



\end{thebibliography}
\end{document}